\documentclass[11pt]{article}
\usepackage{amssymb}
\usepackage{amsmath}
\usepackage{bbm}
\usepackage{amscd}
\usepackage{amsfonts}
\usepackage{graphicx}
\usepackage{color}
\usepackage{cite}

\newtheorem{theorem}{Theorem}[section]

\newtheorem{cor}[theorem]{Corollary}
\newtheorem{lemma}[theorem]{Lemma}
\newtheorem{define}[theorem]{Definition}

\newcommand\btd{\raise 2pt \hbox{$\hat\bigtriangledown$}\hskip 1.5pt}
\newcommand\bt{\raise 2pt \hbox{$\bigtriangledown$}\hskip 1.5pt}

\def\no{\nonumber}

\def\bref#1{(\ref{#1})}
\def\NN{{\mathbb{N}}}
\def\sup{{\rm{sup}}}
\def\R{{\mathbbm{R}}}
\def\mm{{\rm{mm}}}
\def\EE{{\mathcal{E}}}
\def\FF{{\mathcal{F}}}

\newcommand{\qedd}{\hspace*{\fill}$\Box$\medskip}

\def\sat{\hbox{\rm{sat}}}

\def\ST{{\mathcal{S}}}

\def\Y{{\mathbb{Y}}}

\def\Q{{\mathbb{Q}}}

\def\SR{{\mathbf{R}}}
\def\SS{{\mathcal{R}}}

\def\bu{{\mathbf{u}}}
\def\bv{{\mathbf{v}}}

\begin{document}

\title{Matrix Formula of Differential Resultant for
First Order Generic Ordinary Differential Polynomials}
\author{ Zhi-Yong Zhang,  Chun-Ming Yuan,  Xiao-Shan Gao\\
  KLMM,  Academy of Mathematics and Systems Science\\
   Chinese Academy of Sciences}
\date{}
\maketitle

\noindent{\bf Abstract:} In this paper, a matrix representation for
the differential resultant of two generic ordinary differential
polynomials $f_1$ and $f_2$ in the differential indeterminate $y$
with order one and arbitrary degree is given. That is,  a
non-singular matrix is constructed such that its determinant
contains the differential resultant as a factor. Furthermore,  the
algebraic sparse resultant of $f_1, f_2, \delta f_1, \delta f_2$
treated as polynomials in $y, y', y''$ is shown to be a non-zero
multiple of the differential resultant of $f_1, f_2$.
Although very special,  this seems to be the first matrix
representation for a class of nonlinear generic  differential
polynomials.

\vskip5pt\noindent{\bf Keywords:} Matrix formula,  differential
resultant,  sparse resultant,  Macaulay resultant.

\section{Introduction}

Multivariate resultant, which gives a necessary condition for a set
of $n+1$ polynomials in $n$ variables to have common solutions, is
an important tool in elimination theory.
%
%and many algorithms with best complexity bounds for problems such as
%polynomial equation solving and first order quantifier elimination,
%are based on the theory of multivariate resultant
%\cite{canny1,ce2,renegar1}.
%
One of the major issues in the resultant theory is to give a matrix
representation for the resultant,  which allows fast computation of
the resultant using existing methods of determinant computation.
By a matrix representation of the resultant,  we mean a non-singular
square matrix whose determinant contains the resultant as a factor.
There exist stronger forms of matrix representations. For instance,
in the case of two univariate polynomials in one variable,  there
exist matrix formulae named after Sylvester and
B$\acute{\text{e}}$zout,  whose determinants equal the resultant.
Unfortunately,  such determinant formulae do not generally exist for
multivariate resultants.
Macaulay showed that the multivariate resultant can be represented
as a ratio of two determinants of certain Macaulay matrixes
\cite{Ma-1916}. D'Andrea established a similar result for the
multivariate sparse resultant \cite{dandrea1} based on the
pioneering work on sparse resultant \cite{gel,stu1,ce1}.
This paper will study matrix representations for differential
resultants.

Using the analogue between ordinary differential operators and
univariate polynomials,  the differential resultant for two linear
ordinary differential operators was studied by Berkovich and
Tsirulik \cite{berkovich} using Sylvester style matrices.
The subresultant theory was first studied by Chardin \cite{chardin1}
for two differential operators and then by Li \cite{zmli} and Hong
\cite{hhong} for the more general Ore polynomials.

For nonlinear differential polynomials,  the differential resultant
is more difficult to define and study.
The differential resultant for two nonlinear differential
polynomials in one variable was defined by Ritt in
\cite[p.47]{ritt0}.
In \cite[p.46]{handbook},  Zwillinger proposed to define the
differential resultant of two differential polynomials as the
determinant of a matrix following the idea of algebraic multivariate
resultant,  but did not give details.
General differential resultants were defined by Carr\`a Ferro using
Macaulay's definition of algebraic resultants \cite{dres1}. But, the
treatment in \cite{dres1} is not complete,  as will be shown in
Section 2.2 of this paper.
In \cite{yang-dixon},  Yang,  Zeng,  and Zhang used the idea of
algebraic Dixon resultant to compute the differential resultant.
Although very efficient, this approach is not complete and does not
provide a matrix representation for the differential resultant.
Differential resultants for linear ordinary differential polynomials
were studied by Rueda-Sendra \cite{lres1}. In  \cite{lres2},  Rueda
gave a matrix representation for a generic sparse linear system.
In \cite{gao},  the first rigorous definition for the differential
resultant of $n+1$ differential polynomials in $n$ differential
indeterminates was given and its properties were proved.
In \cite{liw1,liw2},  the sparse resultant for differential Laurent
polynomials was defined and a single exponential time algorithm to
compute the sparse resultant was given.
Note that an ideal approach is used in \cite{gao, liw1,liw2},  and
whether the multivariate differential resultant admits a matrix
representation is left as an open issue.

In this paper,  based on the idea of algebraic sparse resultants and
Macaulay resultants,  a matrix representation for the differential
resultant of two generic ordinary differential polynomials $f_1,
f_2$ in the differential indeterminate $y$ with order one and
arbitrary degree is given. The constructed square matrix has entries
equal to the coefficients of $f_1,f_2$, their derivatives, or zero,
whose determinant is a nonzero multiple of the differential
resultant.
Furthermore,  we prove that the sparse resultant of $f_1, f_2,
\delta f_1, \delta f_2$ treated as polynomials in $y, y', y''$ is
not zero and contains the differential resultant of $f_1, f_2$ as a
factor.
Although very special,  this seems to be the first matrix
representation for a class of nonlinear generic differential
polynomials.

The  rest of the paper is organized as follows. In Section 2, the
method of Carr\`a Ferro is briefly introduced and the differential
resultant is defined following \cite{gao}.
In Section 3,  a matrix representation for the differential
resultant of two differential polynomials with order one and
arbitrary degree is given.
In Section 4,   it is shown that the differential resultant can be
computed as a factor of a special algebraic sparse resultant.
In Section 5,  the conclusion is presented and a conjecture is
proposed.

\section{Preliminaries}
To motivate what we do,  we first briefly recall Carr\`a Ferro's
definition for differential resultant and then give a counter
example to show the incompleteness of Carr\`a Ferro's method when
dealing with nonlinear generic differential polynomials. Finally,
definition for differential resultant given in \cite{gao} is
introduced.

\subsection{A sketch of Carr\`a Ferro's definition }

Let $K$ be an ordinary differential field of characteristic zero
with $\delta$ as a derivation operator. $K\{y\}=K[\delta^n y, n\in
\NN]$ is the differential ring of the differential polynomials in
the differential indeterminate $y$ with coefficients in  $K$. Let
$p_1$ (respectively $p_2$)  be a differential polynomial of order
$m$ and degree $d_1$ (respectively of  order $n$ and degree $d_2$ )
in $K\{y\}$.
According to Carr\`a Ferro \cite{dres1}, the differential resultant
of $p_1,  p_2$,  denoted by $\delta \mbox{R}(p_1,  p_2)$,  is
defined to be the Macaulay's algebraic resultant of $m+n+2$
differential polynomials
\begin{eqnarray}
&&\no\mathcal {P}(p_1, p_2)=\{\delta^n p_1, \delta^{n-1} p_1, \dots,
p_1, \delta^m p_2, \delta^{m-1} p_2, \dots, p_2\}
\end{eqnarray}
in the polynomial ring $S_{m+n}=K[y,  \delta y, \dots,
\delta^{m+n}y]$ in $m+n+1$ variables.

Specifically,  let
\begin{eqnarray}
&&\no D=1+(n+1)(d_1-1)+(m+1)(d_2-1), ~~ L={m+n+1+D\choose m+n+1}.
\end{eqnarray}
Let $y_{i}=\delta^iy$ for all $i=0,  1, \dots,  m+n$. For each
$a=(a_0 , \dots,  a_{m+n})\in \NN^{m+n+1}$,  $Y^a=y_0^{a_0}\cdots
y_{m+n}^{a_{m+n}}$ is a power product in $S_{m+n}$. $M_{m+n+1}^{D}$
stands for the set of all power products in $S_{m+n}$ of degree less
than or equal to $D$. Obviously,  the cardinality of $M_{m+n+1}^{D}$
equals $L$. In a similar way it is possible to define
$M_{m+n+1}^{D-d_1}$ which has $L_1={m+n+1+D-d_1\choose m+n+1}$
monomials,  and $M_{m+n+1}^{D-d_2}$ which has
$L_2={m+n+1+D-d_2\choose m+n+1}$ monomials. The monomials in
$M_{m+n+1}^{D}, M_{m+n+1}^{D-d_1}$ and $M_{m+n+1}^{D-d_2}$ are
totally ordered using first the degree and then the lexicographic
order derived from $y_0<y_1<\dots<y_{m+n}$.

\begin{define}\label{def-21}
The $(((n+1)L_1+(m+1)L_2)\times L)$-matrix
\begin{eqnarray}
\no M(\delta,  n,  m)=M(\delta^n p_1, \dots, \delta p_1,  p_1,
\delta^m p_2 , \dots, \delta p_2,  p_2),
\end{eqnarray}
is defined in the following way: for each $i$ such that $(
j-1)L_1<i\leq jL_1$ the coefficients of the polynomial
$Y^a\delta^{n+1-j}p_1$ are the entries of the $i$-th row for each
$Y^a\in M_{m+n+1}^{D-d_1}$ and each $j=1, \dots, n+1$,  while for
each $i$ such that
\begin{eqnarray}
\no (n+1)L_1 +(j-n-2)L_2<i\leq (n+1)L_1+(j-n-1)L_2
\end{eqnarray}
the coefficients of the polynomial $Y^a\delta^{m+n+2-j}p_2$ are the
entries of the $i$-th row for each $Y^a\in M_{m+n+1}^{D-d_2}$ and
each $j=n+2, \dots, m+n+2$,  that are written with respect to the
power products in $M_{m+n+1}^{D}$ in decreasing order.
\end{define}

\begin{define}\label{def-22}
The differential resultant of $p_1$ and $p_2$ is defined to be
 \begin{eqnarray}
% &&\no \delta\text{R}(p_1, p_2)=\text{R}\{\delta^n p_1, \delta^{n-1}
% p_1, \dots, p_1, \delta^m p_2, \delta^{m-1}p_2, \dots, p_2\}\\
 &&\no\hspace{2.28cm}\gcd(\mbox{\rm{det}}(P): P ~\mbox{is an}~ (L\times L)\mbox{-submatrix of}~ M(\delta,  n,  m)).
\end{eqnarray}
\end{define}

\subsection{A counter example and definition of differential resultant}
 In this subsection,  we use Carr\`a Ferro's method \cite{dres1} to
construct matrix formula of two nonlinear generic ordinary
differential polynomials with order one and degree two,
\begin{eqnarray}\label{example}
&&\no g_1=a_0 y_1^2+a_1 y_1y+a_2y^2+a_3y_1+a_4y+a_5, \\
&& g_2=b_0 y_1^2+b_1 y_1y+b_2y^2+b_3y_1+b_4y+b_5,
\end{eqnarray}
where,  hereinafter,  $y_1=\delta y$ and $a_i, b_i$ with $i=0,
\dots, 5$ are generic differential indeterminates.

For differential polynomials in \bref{example},  we have $d_1=d_2=2,
m=n=1$,
 $D=5, L=56, $ and $L_1=L_2=20$. The set of column monomials is
\begin{eqnarray}
\no M_3^5=\{y_2^5, y_2^4B_3^1, y_2^3B_3^2, y_2^2B_3^3, y_2B_3^4,
B_3^5\},
\end{eqnarray}
where,  and throughout the paper,  $B=\{1, y, y_1, y_2\}$ and
$B_i^j$ denotes all monomials of total degree less than or equal to
$j$ in the first $i$ elements of $B$. For example, $B_2^2=\{1, y,
y^2\}$ and $B_3^2=\{1, y, y_1, y^2, yy_1, y_1^2\}$. Note that the
monomials of $M_3^3$ are $M_3^3=\{y_2^3, y_2^2B_3^1, y_2B_3^2,
B_3^3\}=B^3_4$.

According to Definition \ref{def-21},  $M(\delta, 1, 1)$ is an
$80\times 56$ matrix  \begin{center}
\begin{minipage}{7.3cm}
\begin{minipage}{1cm} %b ottom, t op.
$\begin{array}{c}\\
\\
M_3^3 \delta g_1\Bigg\{\\
 \\
M_3^3 g_1\Bigg\{\\
\\
M_3^3 \delta g_2\Bigg\{\\
\\
M_3^3 g_2\Bigg\{\\
 \\
 \end{array}$
\end{minipage}
\begin{minipage}{9.3cm}
\begin{center}
 \hspace{-0.2cm}$y_2^5$ \hspace{0.3cm}$y_2^4y_1$
\hspace{0.3cm}$\dots$\hspace{0.4cm}$y_2^3$
 \hspace{0.3cm}$\dots$\hspace{0.9cm}$y$\hspace{0.7cm}$1$
\vspace{0.1cm}

$\left(
\begin{array}{ccccccc}
0&d_1a_0&\dots&\delta a_5& \dots&0&0\\
  & &&\dots\dots&&&\\
 0 &0& \dots&0&\dots&\delta a_4&\delta a_5\\
0& 0&\dots&a_0&\dots&0&0\\
 & &&\dots\dots&&&\\
0& 0&\dots&0&\dots&a_4&a_5\\0&d_2b_0&\dots&\delta b_5& \dots&0&0\\
  & &&\dots\dots&&&\\
 0 &0& \dots&0&\dots&\delta b_4&\delta b_5\\
0& 0&\dots&b_0&\dots&0&0\\
 & &&\dots\dots&&&\\
0& 0&\dots&0&\dots&b_4&b_5
 \end{array}
\right)$
\end{center}\end{minipage}\end{minipage}\end{center}

Obviously,  the entries of the first column are all zero in
$M(\delta, 1, 1)$,  since the monomial $y^5_2$ never appears in any
row polynomial $Y\ast f$,  where the monomial $Y\in M_3^3$ and $f\in
\{\delta g_1, g_1, \delta g_2, g_2\}$. Consequently,  the
differential resultant of $g_1, g_2$ is identically zero according
to this definition.

Actually,  the differential resultant is defined using an ideal
approach for two generic differential polynomials in one
differential indeterminate in \cite{ritt0} and $n+1$ generic
differential polynomials in $n$ differential indeterminates in
\cite{gao}.
$f$ is said to be a {\em generic differential polynomial} in
differential indeterminates $\Y=\{y_1, \dots, y_n\}$ with order $s$
and degree $m$ if $f$ contains all the monomials of degree up to $m$
in $y_1, \dots, y_n$ and their derivatives up to order $s$.
Furthermore,  the coefficients of $f$ are also differential
indeterminates. For instance, $g_1$ and $g_2$ in \bref{example} are
two generic differential polynomials.

\begin{theorem}[\cite{gao}]\label{th-dr1}
Let $p_0,p_1, \ldots, p_n$ be generic differential polynomials with
order $s_i$ and coefficient sets $\bu_i$ respectively. Then $[p_0,
p_1, \ldots, p_n]$ is a prime differential ideal in $\Q\{\Y, \bu_0,
$ $\ldots, \bu_n\}$. And
 \begin{equation}\label{eq-dr1}
 [p_0,p_1, \ldots, p_n]\cap\Q\{\bu_{0}, \ldots, \bu_{n}\} =
  \sat(\SR(\bu_{0}, \ldots, \bu_{n})).\end{equation}
is a prime differential ideal of codimension one, where $\SR$ is
defined to be the differential sparse resultant of $p_0,p_1, \ldots,
p_n$, which has the following properties

a) $\SR(\bu_{0}, \bu_{1}, \ldots, \bu_{n})$ is an irreducible
polynomial and differentially homogeneous in each $\bu_i$.

b)  $\SR(\bu_{0}, \bu_{1}, \ldots, \bu_{n})$ is of order $h_i=s-s_i$
in $\bu_i$ $(i=0, \ldots, n)$ with $s=\sum_{l=0}^n s_l$.

c) The differential resultant can be written as a linear combination
of $p_i$ and the derivatives of $p_i$ up to the order $s-s_i$.
Precisely,  we have
 $$\SR(\bu_{0}, \bu_{1}, \ldots, \bu_{n}) = \sum_{i=0}^n\sum_{j=0}^{s-s_i}
 h_{ij} p_i^{(j)}$$
\end{theorem}
where $h_{ij}\in
\Q[\Y,\Y^{(1)}\ldots,\Y^{(s)},\bu_0^{(s-s_0)},\ldots,\bu_n^{(s-s_n)}]$.

%\textbf{Definition 2.3} \cite{gao} Let $f_i(i=0, 1, \dots, n)$ be
%generic differential polynomials in $n$ differential variables
%$y_1, \dots, y_n$ with order $s_i$,  degree $m_i$ respectively.
%%
%Let $\mathcal {I}=[f_0, f_1, \ldots, f_n]$ be the differential ideal in
%$K\{y_i, u_0, u_1, \dots, u_n\}$,  where
%$u_i=(u_{i0}, \dots, u_{i\alpha_i})$ are the coefficient sets of $f_i$
%and $\alpha_i$ denotes the number of terms in $f_i$.
%%
%Then  $\mathcal {I}\bigcap K\{u_1, \dots, u_n\}$ is of codimension one
%and can be written as the form $\mbox{sat}(R)$,  where $R$ is an
%irreducible differential polynomial in $u_0, u_1, \dots, u_n$ and is
%defined as the {\em differential resultant} for the differential
%polynomials $f_0, f_1, \ldots, f_n$. Furthermore,  the order of $R$ in
%each $u_i$ is $\sum_{i=0}^n s_i - s_i$.
%
%It is easy to see that with Carr\`a Ferro's method,
%the differential resultant is always zero for differential
%polynomials with positive orders and degree larger than one. In this
%paper,  we present a matrix formula of differential resultant for two
%generic differential polynomials with order one and arbitrary
%degrees.

\section{Matrix formula for differential polynomials}

In this section,  we will give a matrix representation for the
following generic differential polynomials in $y$
\begin{eqnarray}
&&f_1=a_{y_1^{d_1}} y_1^{d_1}+a_{y_1^{d_1-1}y} y_1^{d_1-1}y+\dots+a_0, \label{poly1}\\
&&
 f_2=b_{y_1^{d_2}}y_1^{d_2}+b_{y_1^{d_2-1}y}y_1^{d_2-1}y+\dots+b_0, \no
\end{eqnarray}
where $y_1 = \delta y$,  $1\leq d_1\leq d_2$,  and $a_{y_1^{d_1}},
\dots, a_0, b_{y_1^{d_2}}, \dots, b_0$ are differential
indeterminates.

\subsection{Matrix construction}
In this subsection,  we show that when choosing a proper column
monomial set,  a square matrix can be constructed following
Macaulay's idea \cite{Ma-1916}.

By c) of Theorem \ref{th-dr1},   the differential resultant for
$f_1, f_2$ can be written as a linear combination of $f_1, f_2,
\delta f_1, \delta f_2$ which are treated as polynomials in
variables $y, y_1, y_2=\delta y_1$. So,  we will try to construct a
matrix representation for the differential resultant of $f_1, f_2$
from these four polynomials.

From Section 2,  it is easy to see that the main problem for Carr\`a
Ferro's definition is that the matrix $M(\delta, n, m)$ contains too
many columns. Or equivalently,  the monomial set $M_{m+n+1}^D$ used
to represent the columns is too large.

Consider the monomial set
\begin{equation}\label{eq-E}
\mathcal {E}=B_3^{D}\cup y_2 B_3^{D-1}
\end{equation}
with $D=2d_1+2d_2-3$. We will show that if using $\mathcal {E}$ as
the column monomial set,  a nonsingular square matrix can be
constructed.

Define the {\em main monomial} of polynomials $p_1=\delta f_1,
p_2=\delta f_2, p_3=f_1, p_4=f_2$ to be
 \begin{equation}\label{eq-mm}
 \mbox{mm}(p_1)= y_2\,y_1^{d_1-1}, \mbox{mm}(p_2)=y_1^{d_2},
 \mbox{mm}(p_3)=y^{d_1}, \mbox{mm}(p_4)=1.
 \end{equation}
Then,  we can divide $\mathcal {E}$ into four mutually disjoint sets
$\mathcal {E}=\ST_1\cup \ST_2\cup \ST_3\cup \ST_4$,  where
%\begin{eqnarray}\label{set}
% &&\no \ST_1=\{Y^{\alpha}\in \mathcal {E}: y_2\, y_1^{d_1-1}~\text{divides} ~Y^{\alpha}\}, \\
% && \ST_2=\{Y^{\alpha}\in \mathcal {E}: y_2\, y_1^{d_1-1}~ \text{does not divide}~Y^{\alpha}~\text{but}~ y_1^{d_2}~\text{does}\}, \\\no
% && \ST_3=\{Y^{\alpha}\in \mathcal {E}: y_2\, y_1^{d_1-1}, y_1^{d_2}~
% \text{do not divide}~Y^{\alpha}~\text{but}~
% y^{d_1}~\text{does}\}, \\\no
% && \ST_4=\{Y^{\alpha}\in \mathcal {E}:
% y_2\, y_1^{d_1-1}, y_1^{d_2}, y^{d_1}~ \text{do not
% divide}~Y^{\alpha}\}.
%\end{eqnarray}
%
\begin{eqnarray}\label{set}
 &&\no \ST_1=\{Y^{\alpha}\in \mathcal{E}: \mm(p_1)~\text{divides} ~Y^{\alpha}\}, \\
 && \ST_2=\{Y^{\alpha}\in \mathcal {E}: \mm(p_1)~ \text{does not divide}~Y^{\alpha}~\text{but}~ \mm(p_2)~\text{does}\}, \\\no
 && \ST_3=\{Y^{\alpha}\in \mathcal {E}: \mm(p_1),\mm(p_2)~
 \text{do not divide}~Y^{\alpha}~\text{but}~
 \mm(p_3)~\text{does}\}, \\\no
 && \ST_4=\{Y^{\alpha}\in \mathcal {E}:
 \mm(p_1),\mm(p_2),\mm(p_3)~ \text{do not
 divide}~Y^{\alpha}\}.
\end{eqnarray}
%
%It is easy to see that if $Y^{\alpha}\in \ST_i$ then
%$\mbox{mm}(p_i)$ is a factor of $Y^{\alpha}$.
As a consequence,  we
can write down a system of equations:
\begin{eqnarray}\label{equation}
&&\no Y^{\alpha}/\mbox{mm}(p_1)\ast p_1=0, \hspace{1.83cm}\text{for
}~Y^{\alpha}\in \ST_1, \\\no
&& Y^{\alpha}/\mbox{mm}(p_2)\ast p_2=0,
\hspace{1.83cm}\text{for}~~Y^{\alpha}\in \ST_2, \\
&& Y^{\alpha}/\mbox{mm}(p_3)\ast p_3=0, \hspace{1.83cm}\text{for
}~Y^{\alpha}\in \ST_3, \\\no
&& Y^{\alpha}/\mbox{mm}(p_4)\ast p_4=0, \hspace{1.83cm}\text{for
}~Y^{\alpha}\in \ST_4.
\end{eqnarray}
Observe that the total number of equations is the number of elements
in $\mathcal {E}$ and denoted by $N=(D+1)^2$.

Regarding the monomials in \bref{equation} as unknowns,  we obtain a
system of $N$ linear equations about these monomial unknowns. Denote
the coefficient matrix of the system of linear equations
(\ref{equation}) by $D_{d_1, d_2}$ whose elements are zero or the
coefficients of $f_i$ and $\delta f_i,  i=1, 2$.

Note that the main monomials of the polynomials are not the maximal
monomials in the sense of Macaulay \cite{Ma-1916},  so the monomials
on the left hand side of  (\ref{equation}) may not be contained in
$\mathcal {E}$. Next,  we prove that this  does not occur for our
main monomials.

\begin{lemma}\label{lm-31}
The coefficient matrix $D_{d_1, d_2}$ of system (\ref{equation}) is
square.
\end{lemma}
\emph{Proof:} The coefficient matrix of system (\ref{equation}) has
$N=|\mathcal {E}|$ rows.
In order to prove the lemma,  it suffices to demonstrate that,  for
each $Y^{\alpha}\in \ST_i,i=1,\ldots,4$,  all monomials in
$[Y^{\alpha}/\mbox{mm}(p_i)]\ast p_i$ are contained in $ \mathcal
{E}$. Recall that $\mathcal {E}=B_3^{D}\cup y_2 B_3^{D-1}$. Then by
(\ref{set}),  one has
\begin{eqnarray}\label{case1set}
&&\no \ST_1=
B_3^{D-d_1}\ast\mbox{mm}(p_1)=B_3^{D-d_1}\ast\mbox{mm}(\delta f_1),
\\\no
&& \ST_2=B_3^{D-d_2}\ast\mbox{mm}(p_2)=B_3^{D-d_2}\ast\mbox{mm}(\delta f_2), \\
&& \ST_3=T_1 \ast\mbox{mm}(p_3)=T_1 \ast\mbox{mm}( f_1), \\\no
&& \ST_4=T_2\ast\mbox{mm}(p_4)=T_2\ast\mbox{mm}(f_2),
\end{eqnarray}
where
\begin{eqnarray}\label{T}
&&\no T_1 = \left\{
\Big(\bigcup_{i=0}^{d_2-1}y_1^iB_2^{D-d_1-i}\Big)\bigcup
\Big(y_2\bigcup_{i=0}^{d_1-2}y_1^iB_2^{D-d_1-1-i}\Big)\right\},
\\
&&T_2 =
\left\{\Big(\bigcup_{i=0}^{d_2-1}y_1^iB_2^{d_1-1}\Big)\bigcup \Big(
y_2\bigcup_{i=0}^{d_1-2}y_1^iB_2^{d_1-1}\Big)\right\}.
\end{eqnarray}
Note that the representation of $\ST_2$ in \bref{case1set} is
obtained with the help of the condition $d_1\leq d_2$.

Hence,  the equations (\ref{equation}) become
\begin{eqnarray}
&&\no B_3^{D-d_1}\ast \delta f_1=0,  \nonumber \\
&& B_3^{D-d_2}\ast\delta f_2=0, \label{polyset}\\
&& T_1\ast f_1=0, \nonumber\\
&& T_2 \ast f_2=0.\nonumber
\end{eqnarray}

Since the monomial set of $\delta f_1$ is $B_3^{d_1}\cup y_2\ast
B_3^{d_1-1}$,  the monomial set of $B_3^{D-d_1}\ast \delta f_1$ is
$B_3^{D-d_1}\ast(B_3^{d_1}\cup y_2\ast B_3^{d_1-1}) = B_3^{D}\cup
y_2B_3^{D-1} =\mathcal{E}$. So monomials in the first set of
equations in (\ref{polyset}) are in $\mathcal{E}$.
%
% T_1 \subset B_3^{D-d_1}\cup  y_2 B_3^{D-d_1-1}
%
Since the monomial set of $f_1$ is $B_3^{d_1}=\cup_{l=0}^{d_1} y_1^l
B_2^{d_1-l}$,  the monomial set of $T_1\ast f_1$ is $T_{11}\cup
y_2T_{12}$,  where
 $T_{11} =   \bigcup_{k=0}^{d_1+d_2-1}y_1^k B_2^{D-k}$ and
 $T_{12}= \bigcup_{k=0}^{2d_1-2}y_1^k B_2^{D-k-1}$.
Since $d_1\ge 1$ and $d_2\ge 1$,  we have $d_1+d_2-1\le D =
2d_1+2d_2-3$ and hence $T_{11}\subset B_3^D$.
Since $d_1\ge 1$ and $d_2\ge 1$,  we have $2d_1-2 \le D-1=
2d_1+2d_2-4$ and hence $T_{12} \subset B_3^{D-1}$. As a consequence,
$T_{11}\cup y_2T_{12}\subset \mathcal{E}$ and the monomials in the
third set of equations in (\ref{polyset}) are in $\mathcal{E}$.
Other cases can be proved similarly.
%
%\begin{eqnarray}\label{polyset11}
%\no &&
%\left[\Big(\bigcup_{i=0}^{d_2-1}y_1^iB_2^{D-d_1-i}\Big)\bigcup
%\Big(y_2\bigcup_{i=0}^{d_1-2}y_1^iB_2^{D-d_1-1-i}\Big)\right]\ast
%B_3^{d_1} , \\\no &&
%\left[\Big(\bigcup_{i=0}^{d_2-1}y_1^iB_2^{d_1-1}\Big)\bigcup
%\Big(y_2\bigcup_{i=0}^{d_1-2}y_1^iB_2^{d_1-1}\Big)\right]\ast
%B_3^{d_2},
%\end{eqnarray}
Thus all monomials in the left hand side of (\ref{polyset}) are in
$\mathcal {E}$. This proves the lemma. \qedd

It is worthy to say that,  due to the decrease of the number of
monomials in $\mathcal {E}$ compared with the method by Carr\`a
Ferro,  the size of the matrix $D_{d_1, d_2}$ decreases
significantly.

\subsection{Matrix representation for differential resultant}
In this section,  we show that $\mbox{det}(D_{d_1, d_2})$ is not
identically equal to zero and contains the differential resultant as
a factor.

\begin{lemma}\label{lm-32}
$\hbox{\rm{det}}(D_{d_1, d_2})$ is not identically equal to zero.
\end{lemma}

\emph{Proof:} It suffices to show that there exists a unique
monomial in the sense that it is different from all other monomials
in the expansion of $\mbox{det}(D_{d_1, d_2})$.

The coefficients of the main monomials in $\delta f_1, \delta f_2,
f_1, f_2$ are respectively
\begin{eqnarray}
 &&\no\delta f_1: ~~a_{y_1^{d_1}}\hspace{2.85cm}\mbox{the
  coefficient of}~ \mbox{mm}(\delta f_1)=y_2y_1^{d_1-1}, \nonumber\\
 && \delta f_2: ~~\delta
b_{y_1^{d_2}}+b_{y_1^{d_2-1}y}\hspace{1.1cm}\mbox{the coefficient
of}~\mbox{mm}(\delta f_2)= y_1^{d_2}, \label{coeff}\\
&&\no f_1~: ~~a_{y^{d_1}}\hspace{2.9cm}\mbox{the coefficient of}~
\mbox{mm}(f_1)=y^{d_1}, \nonumber\\
&& f_2~: ~~b_0\hspace{3.25cm}\mbox{the coefficient of}~
\mbox{mm}(f_2)=1.\nonumber
\end{eqnarray}

We will show that the monomial $(a_{y_1^{d_1}})^{n_1}(\delta
b_{y_1^{d_2}})^{n_2}(a_{y^{d_1}})^{n_3}(b_0)^{n_4}$  is a unique one
by the following four steps,  where $n_i$ is the number of elements
in $\ST_i$ with $i=1, \dots, 4$. From \bref{polyset},  $n_1
=|B_3^{D-d_1}|$, $n_2=|B_3^{D-d_2}|$,  $n_3=|T_1|$,  $n_4=|T_2|$.

1. Observe that,  in $\delta f_1$,  $a_{y^{d_1}}$ only occurs in the
coefficient of $y_1y^{d_1-1}$ with the form $a_{y^{d_1}}+\delta
a_{y_{1}y^{d_1-1}}$. Furthermore,  $\delta a_{y_{1}y^{d_1-1}}$ only
occurs in this term given by $\delta f_1$ and no other places of
$D_{d_1, d_2}$. So using the transformation
\begin{eqnarray}\label{tran}
&& \delta a_{y_{1}y^{d_1-1}}=c_{y_1y^{d_1-1}}-d_1 a_{y^{d_1}},
~~\mbox{with other coefficients unchanged, }
\end{eqnarray}
where $c_{y_1y^{d_1-1}}$ is a new differential indeterminate,
$D_{d_1, d_2}$ is transformed to a new matrix which is singular if
and only if the original one is singular. Still denote the matrix by
$D_{d_1, d_2}$.
%
%we obtain that the coefficients of $\delta f_1$ do not contain
%$a_{y^{d_1}}$. Note that transformation (\ref{tran}) is isomorphic
%and does not affect the non-singularity of matrix $D_{d_1, d_2}$.
%

From \bref{polyset},  for a monomial $M\in T_1$,  $a_{y^{d_1}}$ is
the coefficient of the monomial $My^{d_1}$ in each polynomial
$T_1\ast f_1$ and hence in each corresponding row of $D_{d_1, d_2}$.
Then $a_{y^{d_1}}$ is in different rows and columns of $D_{d_1,
d_2}$,  and this gives the factor $(a_{y^{d_1}})^{n_3}$.
Delete those rows and columns of $D_{d_1, d_2}$ containing
$a_{y^{d_1}}$ and denote the remaining matrix by $D^{(1)}_{d_1,
d_2}$.
From \bref{polyset},  the columns deleted are represented by
monomials $y^{d_1}T_1$. So,  $D^{(1)}_{d_1, d_2}$ is still a square
matrix.

2. Let $M\in B_3^{D-d_1}$. The term $a_{y_1^{d_1}}$ occurs in $M\ast
\delta f_1$ as the coefficient of the monomial $y_2y_1^{d_1-1}M$, or
equivalently it occurs in the columns represented by
$y_2y_1^{d_1-1}M$. This gives the factor $(a_{y_1^{d_1}})^{n_1}$. It
is easy to check that $a_{y_1^{d_1}}$ does not occur in other places
of $D^{(1)}_{d_1, d_2}$.
From the definition for $T_1$ \bref{T},  the columns deleted in case
1 correspond those columns represented by monomials of the form
$y_2^{k_2}y_1^{k_1}y^{d_1}$ where either $k_2=0$ and $k_1<d_2$ or
$k_2=1$  and $k_1 < d_1-1$. Then $\{y^{d_1}T_1
\}\cap\{y_2y_1^{d_1-1}B_3^{D-d_1}\}=\emptyset$,  or equivalently
those columns of $D_{d_1, d_2}$ containing $a_{y_1^{d_1}}$ are still
in $D^{(1)}_{d_1, d_2}$.
Similar to case 1,  one can delete those rows and columns of
$D^{(1)}_{d_1, d_2}$ containing $a_{y_1^{d_1}}$ and denote the
remaining matrix by $D^{(2)}_{d_1, d_2}$ which is still a square
matrix. From \bref{polyset},  the columns deleted are represented by
monomials $y_2y_1^{d_1-1}B_3^{D-d_1}$.

3. At the moment,   $D^{(2)}_{d_1, d_2}$ only contains  coefficients
of $f_2$ and $\delta f_2$. Observe that $b_0$ only occurs in the
rows corresponding to $T_2\ast f_2$,  where $T_2$ is defined in
\bref{T}. Note that $\delta b_0$ instead of $b_0$ occurs in $\delta
f_2$.
Since $\{y^{d_1}T_1\}\cap T_2=\emptyset$ and
$\{y_2y_1^{d_1-1}B_3^{D-d_1}\}\cap T_2=\emptyset$,  the columns of
$D_{d_1, d_2}$ containing $b_0$,  represented by $T_2$,  are not
deleted in case 1 and case 2. Then,  we have the factor
$(b_0)^{n_4}$.
Similarly,  delete those rows and columns of $D^{(2)}_{d_1, d_2}$
containing $b_0$ and denote the remaining matrix by $D^{(3)}_{d_1,
d_2}$ which is still a square matrix. From \bref{polyset},  the
columns deleted are represented by monomials $T_2$.
%
%That is to say,  $b_0$ is unique in each row and each column in
%$D^{(2)}_{d_1, d_2}$ if it occurs.

4. From \bref{polyset},  the rows of $D^{(3)}_{d_1, d_2}$ are from
coefficients of $B_3^{D-d_2}\ast \delta f_2$. The term $\delta
b_{y_1^{d_2}}$ is in the coefficient of the monomial $M\ast
y_1^{d_2}$ in $M\ast \delta f_2$ for $M\in B_3^{D-d_2}$,  and
$\delta b_{y_1^{d_2}}$  does not occur in other places of $M\ast
\delta f_2$.
Furthermore,  since $\{y^{d_1}T_1\}\cap
\{y_1^{d_2}B_3^{D-d_2}\}=\emptyset$,
$\{y_2y_1^{d_1-1}B_3^{D-d_1}\}\cap
\{y_1^{d_2}B_3^{D-d_2}\}=\emptyset$,  and $T_2\cap
\{y_1^{d_2}B_3^{D-d_2}\}\}=\emptyset$,   the columns containing the
term $\delta b_{y_1^{d_2}}$ are not deleted in the first three
cases. Then,  we have the factor $(\delta b_{y_1^{d_2}})^{n_2}$.
%
%Thus $\delta b_{y_1^{d_2}}$ occurs in each row and each column of
%$D^{(3)}_{d_1, d_2}$.

Following the above procedures step by step,  the coefficients of
choosing main monomials of the polynomials $f_1,  f_2,  \delta f_1,
\delta f_2$ occur in each row and each column of $D_{d_1, d_2}$ and
only once,  and the monomial
$\big(a_{y_1^{d_1}}\big)^{n_1}\big(\delta
b_{y_1^{d_2}}\big)^{n_2}\big(a_{y^{d_1}}\big)^{n_3}\big(b_0\big)^{n_4}$
is a unique one in the expansion of the determinant of $D_{d_1,
d_2}$. So the lemma follows. \qedd

Note that the selection of main monomials in above algorithm is not
unique,  thus there may exist other ways to construct matrix formula
for system (\ref{poly1}).

\begin{cor}\label{cor-31}
Following the above notations,  for any $Y^{\alpha}\in \ST_i$,   if
all monomials of $[Y^{\alpha}/\text{mm}(p_j)]*p_j$ are contained in
$ \mathcal {E} (j\neq i)$,  then the rearranged matrix,  which is
obtained by replacing the row polynomials
$[Y^{\alpha}/\text{mm}(p_i)]*p_i$ by
$[Y^{\alpha}/\text{mm}(p_j)]*p_j$,  is not identically equal to
zero.
\end{cor}

Corollary \ref{cor-31} follows from the fact that the proof of Lemma
\ref{lm-32} is independent of the number of elements in $\ST_i$ as
long as the main monomials are the same.
%
%As will be demonstrated in Section 4,  the rearranged matrix is
%identical to some algebraic sparse resultant matrix of the system
%$\{f_1, f_2, \delta f_1, \delta f_2\}$ after selecting proper
%perturbed vector and lifted functions.

%\subsection{Properties of the matrix representation}

%Next,  we use the matrix formula of differential resultant of system
%(\ref{poly1}) to show some properties.

The relation between $\mbox{det}(D_{d_1, d_2})$ and differential
resultant of $f_1, f_2$,  denoted by $\SR$,  is stated as the
following theorem.

\begin{theorem}\label{th-31}
$\mbox{\rm{det}}(D_{d_1, d_2})$ is a nonzero multiple of $\SR$.
\end{theorem}
\emph{Proof.} From Lemma \ref{lm-32}, $\det(D_{d_1, d_2})$ is
nonzero. In the matrix $D_{d_1, d_2}$, multiply a column monomial
$M\ne1$ in $\EE$ to the corresponding column and add the result to
the constant column corresponding to the monomial $1$. Then the
constant column becomes $Y^{\alpha}\ast p_i$ with $p_1= \delta f_1,
p_2=\delta f_2, p_3= f_1,  p_4 =f_2$ and $Y^{\alpha} \in
\ST_i/\mbox{mm}(p_i)$, $i=1, \dots, 4$. Since a determinant is
multi-linear on the columns, expanding the matrix by the constant
column,  we obtain
\begin{eqnarray}\label{iden}
&&\mbox{det}(D_{d_1, d_2})= h_1\,  f_1+ h_2\, \delta  f_1+ h_3\,
f_2+ h_4\, \delta f_2,
\end{eqnarray}
where $h_j$ are differential polynomials. From \bref{eq-dr1},
$\mbox{det}(D_{d_1, d_2})\in \mbox{sat}(\SR)$. On the other hand,
from Theorem \ref{th-dr1},  $\SR$ is irreducible and the order of
$\SR$  about the coefficients of $f_1, f_2$ is one. Therefore, $\SR$
must divide $\mbox{det}(D_{d_1, d_2})$.\qedd

From Theorem \ref{th-31},  we can easily deduce a degree bound $N
=4(d_1+d_2-1)^2$ for the differential resultant of $f_1$ and $f_2$.
The main advantage to represent the differential resultant as a
factor of the determinant of a matrix is that we can use fast
algorithms of matrix computation to compute the differential
resultant as did in the algebraic case \cite{ce2}.

Suppose that $\mbox{det}(D_{d_1, d_2})$ is expanded as a polynomial.
Then the differential resultant can be found by the following
result.
\begin{cor}\label{cor-32}
Suppose $\mbox{\rm{det}}(D_{d_1, d_2})=\prod_{i=1}^s P_i^{e_i}$ is
an irreducible factorization of $\mbox{det}(D_{d_1, d_2})$ in
$\Q[C_{f_1},C_{f_2}]$, where $C_{f_i},i=1,2$ are the sets of
coefficients of $f_i$. Then there exists a unique factor, say $P_1$,
which is in $[f_1,f_2]$ and is the differential resultant of $f_1$
and $f_2$.
\end{cor}
\emph{Proof.} From c) of Theorem \ref{th-dr1} and Theorem
\ref{th-31}, $\SR\in [f_1,f_2]$ and is an irreducible factor of
$\mbox{det}(D_{d_1, d_2})$.  Suppose $\mbox{det}(D_{d_1, d_2})$
contains another factor, say $P_2$, which is also in $[f_1,f_2]$.
Then $P_2\in \mbox{sat}(\SR)$ by \bref{eq-dr1}. Since $\SR$ is
irreducible with order one and $P_2$ is of order no more than one,
$P_2$ must equal $\SR$, which contradicts to the hypothesis.\qedd

\subsection{Example (\ref{example}) revisited}
\label{sec-ex1}

In this section,  we apply the method just proposed to construct a
matrix representation for the differential resultant of the system
(\ref{example}).

Following the method given in the proceeding section, for system
(\ref{example}), we have $D=2d_1+2d_2-3=5$ and select the main
monomials of $\delta g_1, \delta g_2, g_1, g_2$ are $y_2y_1, y_1^2,
y^2, 1$ respectively. Then $\mathcal {E}=y_2B_3^4  \cup B_3^5$ is
divided into the following four disjoint sets
\begin{eqnarray}\label{exampleset}
&&\no \ST_1=y_2y_1 B_3^3, \\\no && \ST_2=y_1^2 B_3^3, \\&&
\ST_3=y^2\left[B_2^3\cup y_1B_2^2\cup y_2B_2^2\right], \\\no &&
\ST_4=B_2^{1}\cup y_1B_2^1\cup y_2B_2^{1}.
\end{eqnarray}

Using (\ref{equation}) and regarding the monomials in $\mathcal {E}$
as variables,  we obtain the matrix $D_{2, 2}$,  which is a
$36\times 36$ square matrix in the following form. \vspace{0.2cm}

\begin{minipage}{10.3cm}
\begin{minipage}{4cm} %b ottom, t op.
$\begin{array}{r}\\
\\
B_3^3 \delta g_1\Bigg\{\\
 \\
(B_2^3\cup y_1B_2^2\cup y_2B_2^2) g_1\Bigg\{\\
\\
B_3^3 \delta g_2\Bigg\{\\
\\
(B_2^{1}\cup y_1B_2^1\cup y_2B_2^{1}) g_2\Bigg\{\\
 \\
 \end{array}$
\end{minipage}
\begin{minipage}{9.3cm}
\begin{center}
 \hspace{-0.2cm}$y_2y_1^4$ \hspace{0.15cm}$y_2y_1^3y$ \hspace{0.05cm}$\dots$ \hspace{0.05cm}$y_2y_1^2y^2$
 \hspace{0.1cm}$\dots$\hspace{0.45cm}$y$\hspace{0.6cm}$1$
\vspace{0.2cm}

 $\left(
\begin{array}{ccccccc}
2a_0&a_1&\dots&0& \dots&0&0\\
 & &&\dots\dots&&&\\
 0 &0& \dots&0&\dots&\delta a_4&\delta a_5\\
0& 0&\dots&a_0&\dots&0&0\\
 & &&\dots\dots&&&\\
0& 0&\dots&0&\dots&a_4&a_5\\2b_0&b_1&\dots&0& \dots&0&0\\
 & &&\dots\dots&&&\\
 0 &0& \dots&0&\dots&\delta b_4&\delta b_5\\
0& 0&\dots&b_0&\dots&0&0\\
 & &&\dots\dots&&&\\
0& 0&\dots&0&\dots&b_4&b_5
 \end{array}
\right)$
\end{center}\end{minipage}\end{minipage}\vspace{0.2cm}

As shown in the proof of Lemma \ref{lm-32},
$(a_0)^{10}(a_2)^{10}(b_5)^{6}(\delta b_0)^{10}$ is a unique
monomial in the expansion of the determinant of $D_{2, 2}$. Hence,
the differential resultant of $g_1$ and $g_2$ is a factor of
$\mbox{det}(D_{2, 2})$. Note that in Carr\`a Ferro's construction
for $g_1, g_2$, $M(\delta, 1, 1)$ is an $80\times 56$ matrix,  which
is larger than $D_{2, 2}$.

In particular, suppose $a_0=b_0=1$ and $a_i,b_i$ are differential
constants, i.e, $\delta a_i=\delta b_i=0$, $i=1,\dots, 5$.   Then
$D_{2, 2}$ can be expanded as a polynomial and the differential
resultant of $g_1,g_2$ can be found with Corollary \ref{cor-32},
which is a polynomial of degree 12 and contains 3210 terms. This is
the same as the result obtained in \cite{yang-dixon}.

\section{Differential resultant as the algebraic sparse resultant}
\label{sec-s} In this section,  we show that differential resultant
of $f_1$ and $f_2$ is a factor of the algebraic sparse resultant of
the system $\{f_1, f_2, \delta f_1, \delta f_2\}$.

\subsection{Results about algebraic sparse resultant}
\label{sec-s1} In this subsection, notions of algebraic sparse
resultants are introduced. Detailed can be found in \cite{ce2,
cox-1998, gel, stu1}.

A set $S$ in $\mathbbm{R}^n$ is said to be convex if it contains the
line segment connecting any two points in $S$. If a set is not
itself convex,  its convex hull is the smallest convex set
containing it and denoted by Conv($S$).
A set $V=\{a_1, \ldots, a_m\}$ is called a vertex set of a convex
set $Q$ if each point $q\in Q$ can be expressed as
\begin{eqnarray*}
q=\sum_{j=1}^{m}\lambda_{j}a_{j},  ~~~~\text{with}~
\sum_{j=1}^{m}\lambda_{j}=1~ \text{and}~\lambda_{j}\geq 0,
\end{eqnarray*}
and each $a_{j}$ is called a \emph{vertex} of $Q$.

Consider $n+1$ generic sparse polynomials in the algebraic
indeterminates $x_1, \dots, x_n$:
$$p_i = u_{i0} + u_{i1} M_{i1}+ \cdots + u_{il_i} M_{il_i},  ~~~i=1, \ldots, n+1, $$
where $u_{ij}$ are indeterminates and $M_{ik}=\prod_{s=1}^n
x_s^{e^{ik_s}}$ are monomials in $\Q[x_1, \ldots, x_n]$ with
exponent vectors $a_{ik} = (e^{ik_1}, \ldots,
e^{ik_n})\in\mathbbm{Z}^n$. Note that we assume each $p_i$ contains
a constant term $u_{i0}$.
For $a = (e^{1}, \ldots, e^{n})\in\mathbbm{Z}^n$,  the corresponding
monomial is denoted as $M(a) =\prod_{s=1}^n x_s^{e^{s}}$.

%\begin{define}\label{def-24}\cite{ce2}
The finite set $\mathcal {A}_i\subset \mathbbm{Z}^n$ of all monomial
exponents appearing in $p_i$ is called the \emph{support} of $p_i$,
denoted by supp($p_i$). Its cardinality is $l_i=|\mathcal {A}_i|$.
The {\em Newton polytope} $Q_i\subset \mathbbm{R}^n$ of $p_i$ is the
\emph{convex hull} of $\mathcal {A}_i$,  denoted by $Q_i={\rm
Conv}(\mathcal {A}_i)$. Since $Q_i$ is the convex hull for a finite
set of points,  it must have a vertex set.
For simplicity,  we assume that each $\mathcal {A}_i$ is of
dimension $n$ as did in \cite[p.252]{gel}. Let $\bu$ be the set of
coefficients of $p_i, i=0, \ldots, n$. Then,  the ideal
 \begin{equation}\label{eq-asr}
 (p_1, p_2, \ldots, p_{n+1}) \cap \Q[\bu]=(\SS(\bu))
 \end{equation}
is principal and the generator $\SS$ is defined to be the sparse
resultant of $p_1,  \ldots, p_{n+1}$ \cite[p.252]{gel}. When the
coefficients $\bu$ of $p_i$ are specialized to certain values $\bv$,
the sparse resultant for the specialized polynomials is defined to
be $\SS(\bv)$. The matrix representation of $\SS$ is associated with
the decomposition of the Minkowski sum of the Newton polytopes
$Q_i$.

The \emph{Minkowski sum} of the convex polytopes $Q_i$
\begin{eqnarray}
&&\no Q=Q_1+\dots +Q_{n+1}=\{q_1+\dots+q_{n+1}|q_i\in Q_i\}.
\end{eqnarray}
is still convex and of dimension $n$.

%We first how how to select the column monomials for the sparse
%resultant.
%
Choose sufficiently small numbers $\delta_i>0$ and let $\delta =
(\delta_1, \ldots, \delta_n)\in\R^n$ be a perturbed vector. Then the
points which lie in the interior of the perturbed district
$\EE=\mathbb{Z}^n\cap (Q+\delta)$  are chosen as the {\em column
monomial set} \cite{ce2} to construct the matrix for the sparse
resultant.

%Here,  it is worthy to remark the column monomial set. We directly
%perturb $Q$ to $(Q+\delta)$ and does not consider the specialized
%mixed subdivision of $Q$,  which implies that the dimension of the
%sparse resultant matrix of the system $\{f_1, f_2, \delta f_1,
%\delta f_2\}$ is equal or less than the number of elements of
%$\mathcal {E}$ because some lattice points may not lie in the
%interior of cells of the mixed subdivision \cite{cox-1998}.
%

Choose $n+1$ sufficiently generic linear lifting functions $l_1,
\dots, l_{n+1}\in \mathbb{Z}[x_1, \dots, x_n]$ and define  the
lifted Newton polytopes $\widehat{Q}_i=\{\widehat{q}_i=(q_i,
l_i(q_i)):q_i \in Q_i\}\subset \mathbb{R}^{n+1}$.
Let
$$\widehat{Q}=\sum_{i=1}^{n+1}\widehat{Q}_i\subset\mathbb{R}^{n+1}$$
which is an $(n+1)$-dimensional convex polytope.  The \emph{lower
envelope} of $\widehat{Q}$ with respect to vector $(0, \dots, 0,
1)\in \mathbb{R}^{n+1}$ is the union of all the $n$-dimensional
faces of $\widehat{Q}$,  whose inner normal vector has positive last
component.

Let $\pi: (q_1, \dots, q_{n+1})\mapsto (q_1, \dots, q_{n})$ be the
projection to the first $n$ coordinates from $\mathbb{R}^{n+1}$ to
$\mathbb{R}^{n}$. Then $\pi$ is a one to one map between the lower
envelope of $\widehat{Q}$ and $Q$ \cite{ce2}.
The genericity requirements on $l_i$ assure that every point
$\widehat{q}$ on the lower envelope can be uniquely expressed as
$\widehat{q}=\widehat{q}_1+\dots+\widehat{q}_{n+1}$ with
$\widehat{q}_{i}\in \widehat{Q}_{i}$,  such that the sum of the
projections under $\pi$ of these points leads to a unique sum of
$q=q_1+\dots+q_{n+1} \in Q \subset  \mathbb{R}^n$ with $q_i\in Q_i$,
which is called the \emph{optimal (Minkowski) sum} of $q$.
For $\mathcal{F}_i\subset Q_i$,   $R=\sum_{i=1}^{n+1} \mathcal
{F}_i$ is called an optimal sum,  if each element of $R$ can be
written as a unique optimal sum $\sum_{i=0}^{n} p_i$  for $p_i\in
\mathcal {F}_i$.

A polyhedral \emph{subdivision} of an $n$-dimensional polytope $Q$
consists of finitely many $n$-dimensional polytopes $R_1, \dots,
R_s$,  called the cells of the subdivision,  such that $Q=R_1\cup
\dots\cup R_s$ and for $i\neq j$ and $R_i\cap R_j$ is either empty
or a face of both $R_i$ and $R_j$.
A polyhedral subdivision is called a \emph{mixed subdivision} if
each cell $R_l$ can be written as an optimal sum
$R_l=\sum_{i=1}^{n+1} \mathcal {F}_i$,  where each $\mathcal {F}_i$
is a face of $Q_i$ and $n=\sum_{i=1}^{n+1} \text{dim}(\mathcal
{F}_i)$.
Furthermore,  if $R_j=\sum_{i=1}^{n+1} \mathcal {F}'_i$ is another
cell in the subdivision,  then $R_l\cap R_j=\sum_{i=1}^{n+1}(
\mathcal {F}_i\cap \mathcal {F}'_i)$.
A cell $R_l=\sum_{i=1}^{n+1} \mathcal {F}_i$ is called \emph{mixed}
if dim($\mathcal {F}_i)\leq 1$ for all $i$; otherwise,  it is a
\emph{non-mixed cell}. As a result of $n=\sum_{i=1}^{n+1}
\text{dim}(\mathcal {F}_i)$,  a mixed cell has one unique vertex,
which satisfies $\text{dim}(\mathcal {F}_{i_0})=0$,  while a
non-mixed cell has at least two vertices.

Recall $\delta=(\delta_1, \delta_2, \ldots, \delta_{n})$,  where $0<
\delta_i < 1$. If  $Q=R_1\cup \dots\cup R_s$ is a subdivision of
$Q$,  then $\delta+Q=(\delta+R_1)\cup \dots\cup (\delta+R_s)$ is a
subdivision of $\delta+Q$.

%\textbf{Definition 2.9} \cite{ce2}
%
Let $q\in \mathbb{Z}^n\cap (Q+\delta)$ lie in the interior of a cell
$\delta+\mathcal {F}_1+\dots +\mathcal {F}_{n+1}$  of a mixed
subdivision for $Q+\delta$,  where $\mathcal {F}_i$ is a face of
$Q_i$. The \emph{row content function} of $q$ is defined as the
largest integer such that $\mathcal {F}_i$ is a vertex.
In fact,  all the vertices in the optimal sum of $p$ can be selected
as the row content functions. Hence,  we define \emph{generalized
row content functions} (GRC for brief) as one of the integers,  not
necessary largest,  such that $\mathcal {F}_i$ is a vertex.

Suppose that we have a mixed subdivision of $Q$. With a fixed GRC,
a sparse resultant matrix can be constructed as follows.
For each $i=1, \ldots, n+1$,  define the subset $S_i$ of
$\mathcal{E}$ as follows:
 \begin{equation}\label{eq-si}
 S_i = \{q \in{\mathcal{E}}\, |\, \hbox{GRC}(q)=(i,j_0)\},
 \end{equation}
%that is,  in an optimal sum for $q=q_1+\dots+q_{n+1}$,  $q_i$ is a a
%vertex of $Q_i$. so that
%
where $j_0\in\{1, \dots, m_i\}$,  $m_i$ is the number of vertexes of
$Q_i$, and we obtain a disjoint union for $\mathcal{E}$:

 \begin{equation}\label{eq-si1}
\mathcal{E}=S_1\cup\cdots\cup S_{n+1}.
 \end{equation}
For $q\in S_i$,  let $q=q_1+\dots+q_{n+1}\in Q$ be an optimal sum of
$q$. Then,  $q_i$ is a vertex of $Q_i$ and the corresponding
monomial $M(q_i)$ is called the {\em main monomial} of $p_i$ and
denoted by $\mm(p_i)$,  similar to what we did in Section 3.
%
%We will show that,  for each $q\in S_i$,  the GRC($q$) is unique and
%corresponding main monomials are identical to our choice just below
%\bref{set}.
%
Main monomials have the following important property
\cite[p350]{cox-1998}.

\begin{lemma} \label{lm-mm} If $q\in S_i$,  then the monomials
in $(M(q)/\mm(p_i))p_i$ are contained in ${\mathcal{E}}$.\end{lemma}

Now consider the following equation systems
\begin{equation}\label{eq-mm1}
 (M(q)/\mm(p_i))p_i,~~~~  q\in S_i,  i=1, \ldots, n+1.
 \end{equation}
Treating the monomials in ${\mathcal{E}}$ as variables,  by Lemma
\ref{lm-mm},  the coefficient matrix for the equations in
\bref{eq-mm1} is an $|{\mathcal{E}}|\times |{\mathcal{E}}|$ square
matrix,  called the {\em sparse resultant matrix}. The sparse
resultant of $p_i,  i=1, \ldots, n+1$ is a factor of the determinant
of this matrix.

%
%\subsection{Overview of sparse resultant matrix}
In \cite{ce1, ce2},  Canny and Emiris used linear programming
algorithms  to find the row content functions and to construct
$S_i$. We briefly describe this procedure below.

Now assume $Q_i$ has the vertex set $V_i=\{a_{i1}, \dots, a_{i\,
m_i}\}$. A point $q \in \mathbb{Z}^n\cap (Q+\delta)$ implies that $q
\in \sigma+\delta$ with a cell $\sigma\in Q$. In order to obtain the
generalized row content functions of $q$, we wish to find the
optimal sum of $q-\delta$ in terms of the vertexes in $V_i$.
Introducing variables $\lambda_{ij}, i=1, \ldots, n+1,  j =1,
\ldots, m_i$,   one has
\begin{eqnarray}\label{lcon}
&&q-\delta=\sum_{i=1}^{n+1}q_i=\sum_{i=1}^{n+1}\sum_{j=1}^{m_i}\lambda_{ij}~
a_{ij},  ~~~~\text{with}~ \sum_{j=1}^{m_i}\lambda_{ij}=1~
\text{and}~\lambda_{ij}\geq 0.
\end{eqnarray}
On the other hand,  in order to make the lifted points lie on the
lower envelope of $\widehat{Q}$,  one must force the ``height" of
the listed points minimal,  thus requiring to find $\lambda_{ij}$
such that
\begin{eqnarray}\label{object}
&& \sum_{i=1}^{n+1}\sum_{j=1}^{m_i}\lambda_{ij}~l_i(a_{ij})~~~~
\text{to be minimized}
\end{eqnarray}
under the linear constraint conditions (\ref{lcon}),  where
$l_i(a_{ij})$ is a random linear function in $a_{ij}$.
%
%Obviously,  mixed subdivision is transformed to a linear programming
%problem,  i.e.,  the object function (\ref{object}) with the
%constraint (\ref{lcon}).

For $q\in\EE$,  let $\lambda_{ij}^*$ be an optimal solution for the
linear programming problem \bref{object}. Then
$q-\delta=\sum_{i=1}^{n+1} q_i^*$ where $q_i^* = \sum_{j=1}^{m_i}
\lambda_{ij}^* l_i(a_{ij})$.
$a_{ij_0}^*$ is a vertex of $Q_i$ if and only if there exists a
$j_0$ such that $\lambda_{ij_0}^*=1$ and $\lambda_{ij}^*=0$ for
$j\ne j_0$. In this case,  the generalized row content function of
$q$ is $(i,j_0)$ and $\mm(p_i)$ is $M(a_{ij_0}^*)$.
It is shown that when the lift functions $l_i$ are general enough,
all $S_i$ can be computed in the above way \cite{ce2}.

In order to study the linear programming problem (\ref{object}),  we
need to recall a lemma about the optimality criterion for the
general linear programming problem
\begin{eqnarray}\label{glp}
&&\no \min_{x}  ~~ ~~~~~     z=c^T x\\ && \text{subject to} ~~~~~
Ax=b,  ~~\text{with} ~~l\leq x\leq u,
\end{eqnarray}
where $A$ is an $m\times n$ rectangular matrix,  $b$ is a column
vector of dimension $m$,  $c$ and  $x$ are column vectors of
dimension $n$,  and the superscript $T$ stands for transpose.
In order for the linear programming problem to be meaningful,  the
row rank of $A$ must be less than the column rank of $A$. We thus
can assume $A$ to be row full rank. Let $n_1, \ldots, n_{m}$ be
linear independent columns of $A$. Then the corresponding $x_{n_1},
\ldots, x_{n_m}$ are called {\em basic variables of $x$}. Let $B$ be
the matrix consisting of the $n_1, \ldots, n_{m}$ columns of $A$.
Then $B$ is an $m\times m$ invertible matrix.
Lemma \ref{lm-41} below gives an optimality criterion for the linear
programming problem (\ref{glp}).

\begin{lemma}\label{lm-41}
(\cite{lp}) Let $x_B$ be a basic variables set of $x$,  where $B$ is
the corresponding coefficient matrix of $x_B$.
If the corresponding basic feasible solution $x_B=B^{-1}b\geq 0$ and
the conditions $c_BB^{-1}A-c\leq 0$ hold,  where $c_{B}$ is the row
vector obtained by listing the coefficients of $x_B$ in the object
function,  then an optimal solution for the linear programming
problem (\ref{glp}) can be given as $x_B=B^{-1}b$ and all other
$x_i$ equals zero,  which is called the optimal solution determined
by the basic variables $x_B$.
\end{lemma}

%For the LP problem  (\ref{object}),   given a $q$,  one need to select
%$2n+1$ basic variables and assure that $B$ is of full rank,  and then
%to find basic feasible solutions (\ref{lcon}),  and to check the
%conditions in Lemma 4.1 to find optimal feasible solutions of the
%LP.

\subsection{Algebraic sparse resultant matrix}
\label{sec-s2}In this subsection,  we show that the sparse resultant
for $f_1, f_2, \delta f_1, \delta f_2$ is nonzero and contains the
differential resultant of $f_1$ and $f_2$ as a factor.

For the differential polynomials $f_1$ and $f_2$ given in
\bref{poly1}, consider the  $p_1=\delta f_1, p_2=\delta f_2,
p_3=f_1, p_4=f_2$ as algebraic polynomials in $y, y_1, y_2$.
The monomial sets of $\delta f_1, \delta f_2,f_1, f_2$ are
$B_3^{d_1}\cup y_2\ast B_3^{d_1-1}$, $B_3^{d_2}\cup y_2\ast
B_3^{d_2-1}$, $B_3^{d_1}$, and  $B_3^{d_2}$  respectively.
For convenience,  we will not distinguish a monomial $M$ and its
exponential vector when there exists no confusion.
Then the Newton polytopes for $\delta f_1, \delta f_2, f_1, f_2$ are
respectively,
\begin{eqnarray}
 &&Q_1=\mbox{Conv}(\sup(\delta f_1)) = \mbox{Conv}(B_3^{d_1}\cup y_2\ast B_3^{d_1-1})\subset \R^3, \no\\
 &&Q_2=\mbox{Conv}(\sup(\delta f_2)) = \mbox{Conv}(B_3^{d_2}\cup y_2\ast B_3^{d_2-1})\subset
 \R^3,\no\\
 &&Q_3=\mbox{Conv}(\sup(f_1)) = \mbox{Conv}(B_3^{d_1})\subset \R^3,
 \label{con}\\
 &&Q_4=\mbox{Conv}(\sup(f_2)) = \mbox{Conv}(B_3^{d_2})\subset \R^3,
 \no
\end{eqnarray}
The Newton polytopes $Q_1$ and $Q_3$ are shown in Figure 1 (for
$d_1=5$) while $Q_2$ and $Q_4$ have similar polytopes as $Q_1$ and
$Q_3$ but with different sizes respectively.
\begin{figure}[htp]
\begin{center}
\includegraphics{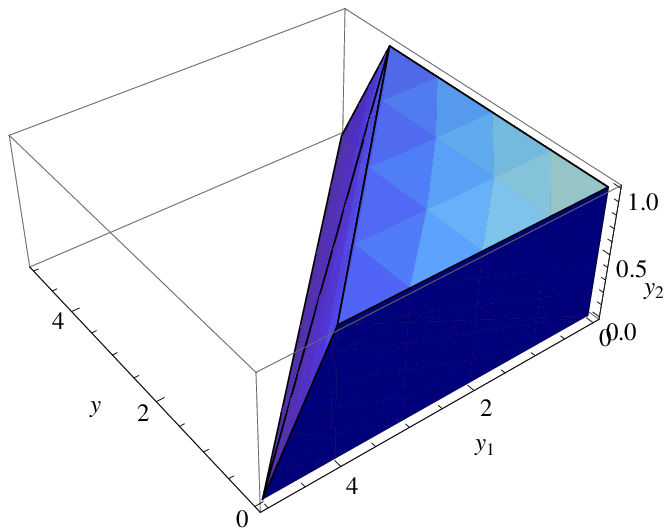}\hspace{1.1cm}
\includegraphics{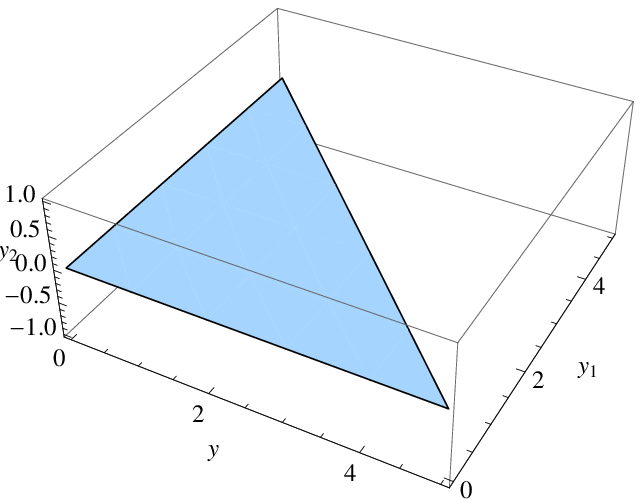}
\end{center}
\caption{The Newton polytopes $Q_1$ and $Q_3$}
\end{figure}

Let the Minkowski sum $Q=Q_1+Q_2+Q_3+Q_4$.
In order to compute the column monomial set,   we choose a perturbed
vector $\delta=(\delta_1, \delta_2, \delta_3)$ with  $0<\delta_i< 1$
with $i=1, 2, 3$. Then the points in $\mathbb{Z}^3\cap(Q+\delta)$ is
easily shown to be $yy_1y_2\mathcal {E}$ where $\EE$ is given in
\bref{eq-E}. Note that using $\EE$ or $yy_1y_2\mathcal {E}$ as the
column monomial set will lead to the same matrix.
%
%In the rest of this section,  $yy_1y_2\mathcal {E}$ will be used.

The vertex sets of $Q_i$,  denoted by $\mbox{V}_{i}$,  are
respectively
\begin{eqnarray*}
&& \mbox{V}_{1}:=\{(0, 0, 0), (0, 0, 1), (0, d_1-1, 1), (0, d_1, 0),
(d_1-1, 0, 1), (d_1, 0, 0)\}, \\\no
&&\mbox{V}_{2}:=\{(0, 0, 0), (0, 0, 1), (0, d_2-1, 1), (0, d_2, 0),
(d_2-1, 0, 1), (d_2, 0, 0)\},\\\no
&&\mbox{V}_{3}:=\{(0, 0, 0), (0, d_1, 0), (d_1, 0, 0)\},\\
&& \mbox{V}_{4}:=\{(0, 0, 0), (0, d_2, 0), (d_2, 0, 0)\}.
\end{eqnarray*}
Let the lifting functions be $l_i=(L_{i1}, L_{i2}, L_{i3}), i=1,
\dots, 4$,  where $L_{ij}$ are parameters to be determined later.
From \bref{object},  the object function of the linear programming
problem to be solved is
{\small \begin{eqnarray}\label{lp}
&& \min_{\lambda_{ij}}(
\lambda_{12}L_{13}+\lambda_{13}[L_{12}(d_1-1)+L_{13}]+\lambda_{14}L_{12}d_1+\lambda_{15}[L_{11}(d_1-1)+L_{13}]
+\lambda_{16}L_{11}d_1\no\\
&&+\lambda_{22}L_{23}+\lambda_{23}[L_{22}(d_2-1)+L_{23}]
+\lambda_{24}L_{22}d_2+\lambda_{25}[L_{21}(d_2-1)+L_{23}]
+\lambda_{26}L_{21}d_2\no\\
&&+\lambda_{32}L_{32}d_1+\lambda_{33}L_{31}d_1\\
&&+\lambda_{42}L_{42}d_2+\lambda_{43}L_{41}d_2)\no
\end{eqnarray}}
under the constraints
\begin{eqnarray}\label{lconfor}
&&\no
A_1=\lambda_{15}(d_1-1)+\lambda_{16}d_1+\lambda_{25}(d_2-1)+\lambda_{26}d_2+\lambda_{33}d_1+\lambda_{43}d_2,
\\\no
&&A_2=\lambda_{13}(d_1-1)+\lambda_{14}d_1+\lambda_{23}(d_2-1)+\lambda_{24}d_2+\lambda_{32}d_1+\lambda_{42}d_2, \\
&&A_3=\lambda_{12}+\lambda_{13} +\lambda_{15}+\lambda_{22}
+\lambda_{23}+\lambda_{25}, \\\no
&&\sum_{j=1}^{m_i}\lambda_{ij}=1,  ~ i=1, \dots, 4, \\
&&\no \lambda_{ij}\geq 0,  i=1, \ldots, 4, j=1, \ldots,
m_i~~\text{with}~
 m_1=m_2=6, m_3=m_4=3,
\end{eqnarray}
where $A_1=\varepsilon_1-\delta_1, A_2=\varepsilon_2-\delta_2,
A_3=\varepsilon_3-\delta_3$ with $(\varepsilon_1, \varepsilon_2,
\varepsilon_3)\in \mathbb{Z}^3\cap (Q+\delta)$.
According to the procedure given in Section \ref{sec-s1},  we need
to solve the linear programming problem (\ref{lp}) for each
$(\varepsilon_1, \varepsilon_2, \varepsilon_3)\in \mathbb{Z}^n\cap
(Q+\delta)$.
Note that $L_{ij}$ are parameters. What we need to do is to show
that there exist $L_{ij}$ such that the solutions of \bref{lp}  make
the corresponding main monomials to be the ones selected by us in
\bref{eq-mm}.
More precisely,  we need to determine $L_{ij}$ such that for each $q
\in \mathbb{Z}^n\cap (Q+\delta)$,  the optimal solution for the
linear programming problem (\ref{lp}) consists one of the following
cases:
\begin{eqnarray*}
&&\lambda_{13}=1\hbox{ implies } \hbox{GRC}(q) =(1,3), \hbox{ the
vertex is } (0, d_1-1, 1),  \hbox{ and } \mm(\delta f_1) =
y_2y_1^{d_1-1},
\\\no
&&\lambda_{24}=1\hbox{ implies } \hbox{GRC}(q) =(2,4), \hbox{ the
vertex is } (0, d_2, 0),  \hbox{ and } \mm(\delta f_2) =
y_1^{d_2},\\
&&\lambda_{33}=1\hbox{ implies } \hbox{GRC}(q) =(3,3), \hbox{ the
vertex is } (d_1, 0, 0),  \hbox{ and } \mm(f_1) = y^{d_1}, \\\no
&&\lambda_{41}=1\hbox{ implies } \hbox{GRC}(q) =(4,1), \hbox{ the
vertex is } (0, 0, 0),  \hbox{ and } \mm(f_2) = 1.
\end{eqnarray*}

The following lemma proves that the above statement is valid.
\begin{lemma}\label{lm-42}
There exist $L_{ij}$ such that the optimal solution of the
corresponding linear programming problem (\ref{lp}) can be chosen
such that the corresponding main monomials for $f_1, f_2, \delta
f_1, \delta f_2$ are
$\text{mm}(f_1)=y^{d_1}$,  $\text{mm}(f_2)=1$,  $\text{mm}(\delta
f_1)=y_2y_1^{d_1-1}$,  $\text{mm}(\delta f_2)=y_1^{d_2}$
respectively and $\EE$ can be written as a disjoint union $\EE=S_1
\cup S_2\cup S_3\cup S_4$,  where $S_i$ is defined in \bref{eq-si}.
\end{lemma}
\emph{Proof.} We write the linear programming problem as the
standard form \bref{glp}. It is easy to see
\begin{eqnarray*}
&&c=(0, L_{13}, (d_1-1) L_{12}+L_{13}, d_1 L_{12}, (d_1-1)L_{11}+L_{13}, d_1 L_{11}, \\
   &&\hskip 0.83truecm 0, L_{23}, (d_2-1) L_{22}+L_{23}, d_2 L_{22}, (d_2-1) L_{21}+L_{23}, d_2L_{21},\\
   &&\hskip 0.83truecm 0, d_1 L_{32}, d_1 L_{31}, 0, d_2 L_{42},
   d_2L_{41}).
\end{eqnarray*}
Let $\delta = (\delta_1, \delta_2, \delta _3)$ be a sufficiently
small vector in sufficiently generic position, the validity of
$\delta$ is analyzed in \cite{ce2}. Then
\begin{center} A=$\left(
\begin{array}{cccccccccccccccccc}
0 & 0 & 0 & 0 & \widetilde{d}_1 & d_1 & 0 & 0 & 0 & 0 & \widetilde{d}_2 & d_2 & 0 & 0 & d_1 & 0 &0  & d_2 \\
0 & 0& \widetilde{d}_1 & d_1 & 0 & 0  & 0 & 0 & \widetilde{d}_2 & d_2& 0 & 0  & 0 & d_1& 0  & 0& d_2 &0   \\
 0 & 1 & 1 & 0 & 1 & 0 & 0 & 1 & 1 & 0 & 1 & 0 & 0 & 0 & 0 & 0 & 0 & 0 \\
 1 & 1 & 1 & 1 & 1 & 1 & 0 & 0 & 0 & 0 & 0 & 0 & 0 & 0 & 0 & 0 & 0 & 0 \\
 0 & 0 & 0 & 0 & 0 & 0 & 1 & 1 & 1 & 1 & 1 & 1 & 0 & 0 & 0 & 0 & 0 & 0 \\
 0 & 0 & 0 & 0 & 0 & 0 & 0 & 0 & 0 & 0 & 0 & 0 & 1 & 1 & 1 & 0 & 0 & 0 \\
 0 & 0 & 0 & 0 & 0 & 0 & 0 & 0 & 0 & 0 & 0 & 0 & 0 & 0 & 0 & 1 & 1 & 1 \
\end{array}
\right), $
\end{center}
where $ \widetilde{d}_1=d_1-1,  \widetilde{d}_2=d_2-1$,  which is a
$7\times 18$ matrix  and $b=(A_1, A_2, A_3, 1, 1, 1, 1)$.
% with
%$(A_1, A_2, A_3)\in \mathbb{Z}^3\cap (Q+\delta)$.
It is easy to see that the rank of $A$ is 7,  since $d_1\ge 1$.

From \bref{eq-E},  we have $\EE=
\mathbb{Z}^3\cap(Q+\delta)=yy_1y_2(B_3^{D}\cup y_2 B_3^{D-1})$,
where $D=2d_1+2d_2-3$.
We will construct a disjoint union  $\EE=S_1\cup S_2\cup S_3\cup
S_4$ like \bref{eq-si1} such that the corresponding main monomials
are respectively $\text{mm}(\delta f_1)=y_2y_1^{d_1-1}$,
$\text{mm}(\delta f_2)=y_1^{d_1}$,$\text{mm}(f_1)=y^{d_1}$,
$\text{mm}(f_2)=1$.

Four cases will be considered.

\emph{Case 1.} We will give the conditions about $L_{ij}$ under
which $\text{mm}(\delta f_1)=y_2y_1^{d_1-1}$,  or equivalently,  the
linear programming problem \bref{lp} has an optimal solution where
$\lambda_{13}=1$.
As a consequence,  $S_1$ will also be constructed.

As shown by Lemma \ref{lm-41},  an optimal solution for a linear
programming problem can be uniquely determined by a set of basic
variables. We will construct the required optimal solutions by
choosing different sets of basic variables. Four sub-cases are
considered.

\hspace{0.3cm}1.1. Selecting basic variables as $\text{vet}_{11}
=\{\lambda_{13},  \lambda_{23}, \lambda_{24},\lambda_{32},
\lambda_{33}, \lambda_{41}, \lambda_{43}\}$ while other variables
are nonbasic variables and equal to zero.
Due the constraint $\lambda_{11}+\lambda_{12}+\dots+\lambda_{16}=1$,
we have $\lambda_{13}=1$.
Then for any such an optimal solution of the linear programming
problem \bref{lp}, in the optimal sum of any element
$q=q_1+q_2+q_3+q_4$,   $q_1 = (0, d_1-1, 1)$ is a vertex of $Q_1$,
$\text{mm}(\delta f_1)=y_2y_1^{d_1-1}$, and the corresponding $q$
belongs to $S_1$ as defined in \bref{eq-si}.

We claim that the basic feasible solutions in $\text{vet}_{11}$ must
be nondegenerate meaning  that all basic variables are positive,
that is,  $x_B=B^{-1}b> 0$.
A mixed cell $R=\sum_{i=1}^4\mathcal {F}_i$,  where $\mathcal {F}_i$
is a face of $Q_i$,  must satisfy the dimension constraint
$\sum_{i=1}^4\dim(\mathcal {F}_i)=3$.
In the cell corresponding to the basic variables $\text{vet}_{11}$,
$\FF_1=(0, d_1-1, 1)$ is a vertex of $Q_1$,  $\FF_4$ is a one
dimensional face of $Q_4$ of the form $\lambda_{41}V_{41}+
\lambda_{43}V_{43}$,  where $V_{41}=(0, 0, 0)$,  $V_{43}=(d_2, 0,
0)$,  and $\lambda_{41}+ \lambda_{43} =1$. $\FF_2$ and $\FF_3$ are
one dimensional faces of $Q_2$ and $Q_3$ respectively.
In order for the dimension constraint $\sum_{i=1}^4\dim(\mathcal
{F}_i)=3$ to be valid,  the claim must be true.
For otherwise,  one of the variables in $\text{vet}_{11}$ must be
zero,  say $\lambda_{41}=0$. Then $\lambda_{43}=1$ and $\FF_4$
becomes a vertex, which implies $\sum_{i=1}^4\dim(\mathcal {F}_i)<
3$, a contradiction.

From Lemma \ref{lm-41},  the coefficient matrix of basic variables
in (\ref{lconfor}) is
\begin{center}
$B_{{11}}=\left(
\begin{array}{ccccccc}
 0 & 0 & 0 & 0 & d_1 & 0 & d_2 \\
 d_1-1 & d_2-1 & d_2 & d_1 & 0 & 0 & 0 \\
 1 & 1 & 0 & 0 & 0 & 0 & 0 \\
 1 & 0 & 0 & 0 & 0 & 0 & 0 \\
 0 & 1 & 1 & 0 & 0 & 0 & 0 \\
 0 & 0 & 0 & 1 & 1 & 0 & 0 \\
 0 & 0 & 0 & 0 & 0 & 1 & 1
\end{array}
\right), $\end{center} with  rank($B_{{11}}$)=7.
For all $(A_1, A_2, A_3)$ and $b=(A_1, A_2, A_3, 1, 1, 1, 1)$,  the
requirement $B_{{11}}^{-1}b> 0$ in Lemma \ref{lm-41} gives
\begin{eqnarray*}
&&\no 1< A_3<2, d_1+d_2< A_2+A_3< 2d_1+d_2,  \\&& 2d_1+d_2<
A_1+A_2+A_3< 2d_1+2d_2.
\end{eqnarray*}
Substituting $A_1=\varepsilon_1-\delta_1,
A_2=\varepsilon_2-\delta_2, A_3=\varepsilon_3-\delta_3$ into the
above inequalities and considering that $(\varepsilon_1,
\varepsilon_2, \varepsilon_3) $ are integer points,  we have
\begin{eqnarray}\label{eq-a11}
&& \varepsilon_3=2, ~~\varepsilon_2=d_1+d_2-1, \dots, 2d_1+d_2-2,\\
&&\no\varepsilon_1+\varepsilon_2=2d_1+d_2-1, \dots, 2d_1+2d_2-2.
\end{eqnarray}

On the other hand,  $c_{B_{_{11}}}=((d_1-1) L_{12}+L_{13},
(d_2-1)L_{22}+L_{23}, d_2L_{22}, d_1L_{32}, d_1 L_{31}, 0,$ $
d_2L_{41})$.
After simplification and rearrangement, the condition
$c_{B_{_{11}}}B_{_{11}}^{-1}A-c\leq 0$ in Lemma \ref{lm-41} becomes
\begin{eqnarray}\label{eq-l11}
&&\no \{L_{12}-L_{11}+L_{31}-L_{32},L_{12}+L_{31}-L_{32}-L_{41},
\\&&\hspace{0.2cm}L_{13}-L_{12}+L_{22}-L_{23}, L_{22}-L_{21}+L_{31}-L_{32},\\
&&\no
\hspace{0.2cm}L_{22}+L_{31}-L_{32}-L_{41},L_{31}-L_{41},L_{32}-L_{31}+L_{41}-L_{42}\}\leq
0
\end{eqnarray}
where,  hereinafter,   $\{w_1, \dots, w_s\}\leq0$ means $w_{i}\leq
0$ for $i=1, \dots, s$.

By Lemma \ref{lm-41},  if  \bref{eq-a11} and \bref{eq-l11}  are
valid, we obtain an optimal solution of the linear programming
problem \bref{lp} which is determined by the basic variables
$\text{vet}_{11}$.
Hence, if \bref{eq-l11} is valid, the corresponding
$q=(\varepsilon_1, \varepsilon_2, \varepsilon_3)$ in and
\bref{eq-a11} are in $S_1$, since in the optimal decomposition of
$q=q_1+q_2+q_3+q_4$,  $q_1 = (0, d_1-1, 1)$ is a vertex.

%By Lemma \ref{lm-41},  if \bref{eq-l11} is true,  \bref{eq-a11} are
%optimal solutions of the linear programming problem \bref{lp} and
%hence the corresponding $(\varepsilon_1, \varepsilon_2,
%\varepsilon_3)$ are in $S_1$.

\hspace{0.3cm}1.2. Similarly,  choosing the basic variables as
$\text{vet}_{12} =\{\lambda_{13}, \lambda_{23},
\lambda_{24},\lambda_{31}, \lambda_{32},  \lambda_{33},
\lambda_{41}\}$, which generates a new basic matrix $B_{{12}}$, and
from $B_{{12}}^{-1}b> 0$,  we obtain
\begin{eqnarray*}
&& 0<A_1, d_1+d_2< A_2+A_3, 1<A_3<2,A_1+A_2+A_3< 2d_1+d_2,
\end{eqnarray*}
which in turn  lead to the following values for $\varepsilon_1,
\varepsilon_2, \varepsilon_3$
\begin{eqnarray}\label{eq-a12}
 &&\varepsilon_3=2, ~~\varepsilon_1=1,\dots, ~~\varepsilon_2=d_1+d_2-1,\dots, \\
 &&\varepsilon_1+\varepsilon_2=d_1+d_2, \dots,  2d_1+d_2-2.\no
\end{eqnarray}
The condition $c_{B_{{12}}}B_{{12}}^{-1}A-c\leq 0$ leads to the
following constraints on $L_{ij},  i=1, \dots, 4, j=1, 2, 3$,
\begin{eqnarray}\label{eq-l12}
 && \no\{L_{12}-L_{11}+L_{31}-L_{32},L_{12}-L_{32},\\
 &&\hspace{0.2cm}L_{13}-L_{12}+L_{22}-L_{23},L_{22}-L_{32},\\
 &&\no\hspace{0.2cm}
 L_{22}-L_{21}+L_{31}-L_{32},L_{31}-L_{41},L_{32}-L_{42}\}\leq 0.
\end{eqnarray}

\hspace{0.3cm}1.3. Similarly,  the basic variables $ \text{vet}_{13}
= \{ \lambda_{13}, \lambda_{23}, \lambda_{24},\lambda_{32},
\lambda_{33}, \lambda_{41},$ $ \lambda_{42}\}$ lead to
\begin{eqnarray} \label{eq-a13}
&& \no\varepsilon_3=2, ~~ \varepsilon_1=1, \dots, d_1, ~~\\
&& \varepsilon_1+\varepsilon_2=2d_1+d_2-1, \dots,  2d_1+2d_2-2,
\end{eqnarray}
and
\begin{eqnarray}\label{eq-l13}
&&\no\{L_{12}-L_{42},L_{12}-L_{11}+L_{31}-L_{32},\\
&&\hspace{0.2cm}L_{13}-L_{12}+L_{22}-L_{23},L_{22}-L_{21}+L_{31}-L_{32},\\
&&\no\hspace{0.2cm}L_{22}-L_{42},L_{31}-L_{32}-L_{41}+L_{42},L_{32}-L_{42}\}\leq
0.
\end{eqnarray}

\hspace{0.3cm}1.4. Similarly, the basic variables $ \text{vet}_{14}
= \{ \lambda_{13}, \lambda_{23}, \lambda_{24},\lambda_{33},
\lambda_{41}, \lambda_{42},$ $\lambda_{43}\}$ lead to
\begin{eqnarray} \label{eq-a14}
&&\no \varepsilon_3=2, ~~ \varepsilon_1=d_1+1, \dots,
~~\varepsilon_2=d_1+d_2-1,\dots,\\
&&\varepsilon_1+\varepsilon_2=2d_1+d_2, \dots,  2d_1+2d_2-2.
\end{eqnarray}
and
\begin{eqnarray}\label{eq-l14}
&&\no\{L_{12}-L_{11}+L_{41}-L_{42},L_{12}-L_{42},\\&&
\hspace{0.2cm}L_{13}-L_{12}+L_{22}-L_{23},L_{22}-L_{21}+L_{41}
   -L_{42},\\&&\no\hspace{0.2cm}L_{22}-L_{42},L_{31}-L_{41},L_{31}-L_{32}-L_{41}+L_{42}\}\leq
0.
\end{eqnarray}

Let $S_1$ be the set $(\varepsilon_1, \varepsilon_2, \varepsilon_3)$
defined in \bref{eq-a11},  \bref{eq-a12}, \bref{eq-a13}, and
\bref{eq-a14}. Then for $\eta\in S_1$ and an optimal sum of $\eta =
q_1+q_2+q_3+q_4$, since $\lambda_{13}=1$,  $q_1$ must be the vertex
$(0,d_1-1,1)$ of $Q_1$. Therefore,  $\text{mm}(\delta
f_1)=y_2y_1^{d_1-1}$.
Of course,  in order for this statement to be valid,  $L_{ij}$ must
satisfy constraints \bref{eq-l11}, \bref{eq-l12}, \bref{eq-l13}, and
\bref{eq-l14}. We will show later that these constraints indeed have
common solutions.

%The two selected basic variables $\text{vet}_{11}$ and
%$\text{vet}_{12}$ for (\ref{lconfor}) make $\lambda_{33}=1$,  which
%corresponds to $\text{mm}(f_1)=y^{d_1}$ and is identical to our
%choices for $f_1$.

The following three cases can be treated similarly,  and we only
list the conditions for $\varepsilon_1, \varepsilon_2,
\varepsilon_3$ while the concrete requirements for $L_{ij}$ are
listed at the end of the proof.

\emph{Case 2.} In order for $\text{mm}(\delta f_2)=y_1^{d_1}$,  we
choose the basic variables
\begin{eqnarray}
&&\no \text{vet}_{21} = \{ \lambda_{13},  \lambda_{14},
\lambda_{24},\lambda_{31},  \lambda_{32}, \lambda_{33},
\lambda_{41}\},
\\\no
&& \text{vet}_{22} = \{ \lambda_{13},  \lambda_{14},
\lambda_{24},\lambda_{33},  \lambda_{41},  \lambda_{42},
\lambda_{43}\},
\\\no
&& \text{vet}_{23} = \{\lambda_{13}, \lambda_{14},
\lambda_{24},\lambda_{32}, \lambda_{33}, \lambda_{41},
\lambda_{43}\},\\\no
&&\text{vet}_{24} = \{\lambda_{13}, \lambda_{14},
\lambda_{24},\lambda_{32}, \lambda_{33}, \lambda_{41},
\lambda_{42}\},\\\no &&
\text{vet}_{25} = \{\lambda_{11}, \lambda_{12}, \lambda_{13},
\lambda_{24},\lambda_{31}, \lambda_{33}, \lambda_{41}\},  \\\no
&&\text{vet}_{26} = \{ \lambda_{13}, \lambda_{14}, \lambda_{15},
\lambda_{24},\lambda_{33}, \lambda_{41},  \lambda_{43}\},\\\no
&&\text{vet}_{27} = \{ \lambda_{12}, \lambda_{13},
\lambda_{14},\lambda_{15}, \lambda_{24}, \lambda_{33},
\lambda_{41}\},
\end{eqnarray}
which lead to the following elements of $S_2$
%\begin{eqnarray}
%&&\no 0 < A_1,  d_1+d_2-1< A_2, A_1+A_2< 2d_1+d_2-1, 0<A_3<1;
%\\\no &&d_1 < A_1,  d_1+d_2-1< A_2,  A_1+A_2< 2d_1+2d_2-1, 0<A_3<1;
%\\\no &&d_1+d_2-1 < A_2< 2d_1+d_2-1,  2d_1+d_2-1< A_1+A_2< 2d_1+2d_2-1,  0<A_3<1;
%\\\no &&0< A_1< d_1,  2d_1+d_2-1< A_1+A_2< 2d_1+2d_2-1, 0<A_3<1;
%\\\no &&0 < A_1< d_1,  d_2< A_2< d_1+d_2-1, 0<A_3<1; \\\no
%&&d_2 < A_2< d_1+d_2-1,  2d_1+d_2-1< A_1+A_2< 2d_1+2d_2-1, 0<A_3<1.
%\end{eqnarray}
\begin{eqnarray}
&&\no \varepsilon_3=1, ~~ 0<\varepsilon_1, ~~d_1+d_2-1<
\varepsilon_2,  \varepsilon_1+\varepsilon_2=d_1+d_2+1, \dots,
2d_1+d_2-1;\\\no&& \varepsilon_3=1, ~~ d_1<\varepsilon_1,
~~d_1+d_2-1< \varepsilon_2, \varepsilon_1+\varepsilon_2=2d_1+d_2+1,
\dots,  2d_1+2d_2-1;\\\no&& \varepsilon_3=1, ~~
\varepsilon_2=d_1+d_2, \dots, 2d_1+d_2-1, ~~
\varepsilon_1+\varepsilon_2=2d_1+d_2, \dots,  2d_1+2d_2-1;\\\no&&
\varepsilon_3=1, ~~ \varepsilon_1=1, \dots, d_1, ~~
\varepsilon_1+\varepsilon_2=2d_1+d_2, \dots,  2d_1+2d_2-1;\\\no&&
\varepsilon_3=1, ~~ \varepsilon_1=1, \dots, d_1, ~~
\varepsilon_2=d_2+1, \dots, d_1+d_2-1;\\\no&& \varepsilon_3=1, ~~
\varepsilon_2=d_2+1, \dots, d_1+d_2-1, ~~ \varepsilon_2=2d_1+d_2,
\dots,  2d_1+2d_2-1;\\\no
&&\varepsilon_3=1, ~~ \varepsilon_1=d_1+1, \dots,
~~\varepsilon_2=d_2+1,\dots, \varepsilon_1+\varepsilon_2=d_1+d_2+2,
\dots,  2d_1+d_2-1.
\end{eqnarray}

\emph{Case 3.} In order for $\text{mm}(f_1)=y_1^{d_1}$, we choose
the basic variables
\begin{eqnarray}
&&\no\text{vet}_{31} =\{ \lambda_{15}, \lambda_{16},  \lambda_{24},
\lambda_{26},
\lambda_{33}, \lambda_{41},  \lambda_{43}\},\\
&&\no\text{vet}_{32} =\{\lambda_{13},  \lambda_{15}, \lambda_{23},
\lambda_{24}, \lambda_{33},  \lambda_{41},  \lambda_{43} \},\\
&&\no\text{vet}_{33} = \{\lambda_{15}, \lambda_{23}, \lambda_{24},
\lambda_{25}, \lambda_{33}, \lambda_{41},\lambda_{43}\},\\
&&\no\text{vet}_{34} = \{\lambda_{12}, \lambda_{13}, \lambda_{15},
\lambda_{23}, \lambda_{24}, \lambda_{33}, \lambda_{41}\},
\end{eqnarray}
which lead to the following results about $\varepsilon_1,
\varepsilon_2, \varepsilon_3$ in $S_3$
%\begin{eqnarray}
%&&\no 0 < A_2< d_2-1,  d_2-1< A_1+A_2< d_1+d_2-1,  A_3<2;
%\\\no &&d_1+d_2-2 < A_2< 2d_1+d_2-2,  2d_1+d_2-2< A_1+A_2< 2d_1+2d_2-2,  A_3=2;
%\\\no &&0 < A_1, d_1+d_2-2< A_2, A_1+A_2< 2d_1+d_2-2, A_3=2; \\\no
%&&0 < A_1< d_1,   2d_1+d_2-2< A_1+A_2< 2d_1+2d_2-2,  A_3=2;\\\no &&0 <
%A_2< d_2-1,  2d_1+d_2-2< A_1+A_2< 2d_1+2d_2-2,  A_3=2.
%\end{eqnarray}
\begin{eqnarray}
&&\no\varepsilon_3=1, ~~ \varepsilon_2=1, \dots, d_2, ~~
\varepsilon_1+\varepsilon_2=2d_1+d_2, \dots,  2d_1+2d_2-1;\\
 &&\varepsilon_3=2, ~~\varepsilon_2=d_2, \dots, d_1+d_2-2,\varepsilon_1+\varepsilon_2=2d_1+d_2-1, \ldots,
 2d_1+2d_2-2;\\
&&\no \varepsilon_3=2, ~~ \varepsilon_2=1, \dots, d_2-1, ~~
\varepsilon_1+\varepsilon_2=2d_1+d_2-1, \dots,  2d_1+2d_2-2;\\
&&\no \varepsilon_3=2, ~~ \varepsilon_1=d_1+1, \dots,
~~\varepsilon_2=d_2,\dots, \varepsilon_1+\varepsilon_2=d_1+d_2+1,
\dots, 2d_1+d_2-2.
\end{eqnarray}

\emph{Case 4.} In order for $\mm(f_2)=1$,  we choose the following
basic variables
\begin{eqnarray}
&&\no \text{vet}_{41} = \{ \lambda_{11}, \lambda_{12}, \lambda_{21},
\lambda_{24},  \lambda_{26},\lambda_{31},  \lambda_{41}\},
\\\no
&& \text{vet}_{42} = \{ \lambda_{11},  \lambda_{12},  \lambda_{24},
\lambda_{26},\lambda_{31},  \lambda_{33},  \lambda_{41}\}, \\\no
&& \text{vet}_{43} = \{\lambda_{11}, \lambda_{12},  \lambda_{15},
\lambda_{24},  \lambda_{26}, \lambda_{33}, \lambda_{41}\},  \\\no
&&\text{vet}_{44} = \{\lambda_{12}, \lambda_{13},  \lambda_{23},
\lambda_{24},\lambda_{31},  \lambda_{33},  \lambda_{41}\}, \\\no
&& \text{vet}_{45} = \{ \lambda_{12}, \lambda_{23},  \lambda_{24},
\lambda_{25},\lambda_{31},  \lambda_{33},  \lambda_{41}\},\\\no
&&\text{vet}_{46} = \{ \lambda_{12}, \lambda_{21},
\lambda_{22},\lambda_{23}, \lambda_{25}, \lambda_{31},
\lambda_{41}\},\\\no
&&\text{vet}_{47} = \{ \lambda_{12}, \lambda_{15},
\lambda_{23},\lambda_{25}, \lambda_{26}, \lambda_{33},
\lambda_{41}\},
\end{eqnarray}
which correspond to the elements in $S_4$ %\begin{eqnarray}
%&&\no 0 < A_1,  0< A_2, A_1+A_2< d_2,  0<A_3<1;
%\\\no && 0< A_2< d_2, d_2< A_1+A_2< d_1+d_2,  0<A_3<1;
%\\\no && 0 < A_2< d_2,  d_1+d_2< A_1+A_2< 2d_1+d_2-1, A_3<1; \\\no
%&&0< A_1< d_1, d_2-1< A_2< d_1+d_2-2, 0<A_3<2;
%\\\no &&d_2< A_1< d_1+d_2-1, A_2=d_2-1,  1<A_3.
%\end{eqnarray}
\begin{eqnarray}
&&\no \varepsilon_3=1, ~~ 0<\varepsilon_1, ~~0< \varepsilon_2, ~~
\varepsilon_1+\varepsilon_2=2, \dots, d_2;
\\\no && \varepsilon_3=1, ~~\varepsilon_2=1, \dots, d_2, ~~\varepsilon_1+\varepsilon_2=d_2+1, \dots, d_1+d_2;
\\\no &&
\varepsilon_3=1, ~~\varepsilon_2=1, \dots, d_2,
~~\varepsilon_1+\varepsilon_2=d_1+d_2+1, \dots, 2d_1+d_2-1;
\\\no &&
\varepsilon_3=2, ~~\varepsilon_1=1, \dots, d_1, ~~\varepsilon_2=d_2,
\dots, d_1+d_2-2;\\\no && \varepsilon_3=2, ~~ \varepsilon_2=1,
\dots, d_2-1, ~~ \varepsilon_1+\varepsilon_2=d_2, \dots,
d_1+d_2-1;\\\no
&&\varepsilon_3=2, ~~ \varepsilon_1=1, \dots,
~~\varepsilon_2=1,\dots,d_2-2,~~ \varepsilon_1+\varepsilon_2=2,
\dots, d_2-1;\\\no
&&\varepsilon_3=1, ~~\varepsilon_2=1,\dots,d_2-1,
\varepsilon_1+\varepsilon_2=d_1+d_2-1, \dots,  2d_1+d_2-2.
\end{eqnarray}

Merge all the  constraints for $L_{ij}$,  we obtain
\begin{eqnarray}\label{lc}
 && L_{11}-L_{12}-L_{21}+L_{22}\leq 0, \no\\
 && L_{13}\leq L_{23}, L_{21}\leq L_{31}\leq L_{11}\leq L_{41}, \\
 && \no L_{22}\leq L_{12}\leq L_{32} \leq L_{42},L_{31}=L_{32}+L_{41}-L_{42}.
\end{eqnarray}
The  solution set for system (\ref{lc}) is nonempty. For example,
$l_1=(7, -4, -5), l_2=(5, -9, 5), l_3=(6, 2, 1), l_4=(8, 4, 7)$,
which will be used for example (\ref{example}),  satisfy the
conditions in (\ref{lc}).

We can also check that $\EE = S_1\cup S_2\cup S_3\cup S_4$ is a
disjoint union for $\EE$. The lemma is proved.\qedd

 We now have the main result of this section.
\begin{theorem}\label{th-2}
The sparse resultant of $f_1,  f_2,  \delta f_1,  \delta f_2$ as
polynomials in $y, y_1, y_2$ is not identically zero and contains
the differential resultant of $f_1$ and $f_2$ as a factor.
\end{theorem}
\noindent{\em Proof.} Note that  $a_0,  b_0,  \delta a_0,  \delta
b_0$,  which are the zero degree terms of $f_1$,  $f_2$,  $\delta
f_1$,  $\delta f_2$ respectively,  are algebraic indeterminates. As
a consequence,
 $$J_1= (f_1,  f_2,  \delta f_1, \delta f_2)$$
is a prime ideal in $\Q[\bu, y, y_1, y_2]$,  where $\bu$ is the set
of the coefficients of $f_1, f_2$ are their first order derivatives.
Let
 $$J_2= J_1\cap\Q[\bu].$$
Then $J_2$ is also a prime ideal.
We claim that
 \begin{equation}\label{eq-j2}
 J_2= (\SR)\end{equation}
where $\SR$ is the differential resultant of $f_1, f_2$.
From c) of Theorem \ref{th-dr1},  $\SR\in J_2$.
Let $T\in J_2$. Then $T\in J_1 \subset [f_1, f_2]$. From
\bref{eq-dr1},  the pseudo remainder of $T$ with respect to $\SR$ is
zero. Also note that the order of $T$ in $a_i,  b_i$ is less than or
equal to $1$. From a) and b) of Theorem \ref{th-dr1},  $\SR$ must be
a factor of $T$,  which proves \bref{eq-j2}.

From Lemma \ref{lm-42},  the main monomials for $f_1, f_2, \delta
f_1,  \delta f_2$ are the same as those used to construct $\ST_1,
\ST_2, \ST_3, \ST_4$ in \bref{eq-mm}. As a consequence, we have
$S_1\subset \ST_1$. For $q\in\ST_1\setminus S_1$, $q$ must be in
some $S_i$, say $q\in S_2$. Then from Lemma \ref{lm-42}, the
monomials in $(M(q)/\mm(\delta f_2))  \delta f_2 $ are contained in
$\EE$. By Corollary \ref{cor-31}, the sparse resultant matrix of
$f_1,  f_2,  \delta f_1,  \delta f_2$ obtained after move $q$ from
$S_2$ to $S_1$ is still nonsingular.
Doing such movements repeatedly will lead to $\ST_1=S_1, \ST_2=S_2,
\ST_3=S_3, \ST_4=S_4$. As a consequence, the sparse resultant is not
identically zero.

%From the construction of the main monomials for $f_1, f_2, \delta
%f_1,  \delta f_2$ and the disjoint union $\EE=S_1\cup S_2\cup
%S_3\cup S_4$ in the proof of Lemma \ref{lm-42},  we see that the
%corresponding sparse resultant matrix  equals the rearranged matrix
%defined in Corollary \ref{cor-31},  which means that the sparse
%resultant $\SS$ for $f_1,  f_2,  \delta f_1,  \delta f_2$ is not
%identically zero.
%
From \bref{eq-asr},  we have $\SS\in J_1$ which implies $\SS\in
J_2$. Since $\SR$ is irreducible,  $\SR$ must be a factor of
$\SS$.\qedd

%Note that there are two definitions \bref{set} and \bref{eq-si} for
%$S_1, S_2, S_3, S_4$,  which are not necessarily the same. However,
%by Corollary \ref{cor-31},  we show that the matrix induced by $S_i$
%defined in \bref{eq-si} is not equal to zero identically and
%contains differential resultant of $f_1$ and $f_2$ as a factor.
%Moreover,  as demonstrated by example (\ref{example}),  the $S_i$
%defined in \bref{set} can be changed to those \bref{eq-si} and hence
%lead to the same matrix.

\subsection{Example (\ref{example}) revisited}

We show how to construct a nonsingular algebraic sparse resultant
matrix of the system $\{g_1, g_2, \delta g_1, \delta g_2\}$, where
$g_1,g_2$ are from \bref{example}.

Using the algorithm for sparse resultant in \cite{ce1, ce2}, we
choose perturbed vector $\delta=(0.01, 0.01, 0.01)$ and the lifting
functions $l_1=(7, -4, -5), l_2=(5, -9, 5), l_3=(6, 2, 1), l_4=(8,
4, 7)$,  where $l_i$ corresponds to $Q_i$ defined in (\ref{con})
with $d_1=d_2=2$. These lift functions satisfy the conditions
\bref{lc}.

By Lemma \ref{lm-42},  the main monomials for $g_1, g_2, \delta g_1,
\delta g_2$ are identical with those given in Section \ref{sec-ex1}.
Let $S_1, S_2, S_3, S_4$ be those constructed as in the proof of
Lemma \ref{lm-42}.
%
%Then, we can check $\ST_1 = S_1\cup \{y_2y_1y^3, y_2y_1y^2\}\cup
%\{y_2y_1y, y_2y_1\}$, $\ST_3 =\{y_2y^2, y_1y^2, y^3, y^2\}\cup S_4$.
After the following changes
\begin{eqnarray}
&&\no\text{move}~ \{y_2y_1y^3, y_2y_1y^2\}~\text{in $S_3$ to $S_1$},
\\\no
&& \text{move}~\{y_2y^2, y_1y^2, y^3, y^2\}~\text{in $S_4$ to
$S_3$},\\\no
&& \text{move}~\{y_2y_1y, y_2y_1\}~ \text{in $S_4$ to $S_1$},
\end{eqnarray}
we have $\ST_i = S_i, i=1,\ldots,4$. Then by Corollary \ref{cor-31},
the sparse resultant matrix constructed with the original
$S_1,S_2,S_3,S_4$ is nonsingular and contains the differential
resultant as a factor.

%Then,  by Lemma \ref{lm-42},  the main monomials for $g_1, g_2,
%\delta g_1, \delta g_2$ are identical with those given in
%\bref{exampleset} and the sparse resultant matrix of the system
%$\{g_1, g_2, \delta g_1, \delta g_2\}$ is identical to the matrix
%$\widehat{D}_{2, 2}$ in Section 3.3.
%
%In addition,  by Corollary 3.3,  the rearranged matrix is not equal
%to zero identically as long as it is square after adjusting the
%column monomials sets $\ST_i$ in (\ref{exampleset}). For example,
%one can adjust $\ST_i$ as follows
%\begin{eqnarray}
%&&\no\text{move}~ \{y_2y_1y^3, y_2y_1y^2\}~\text{in $S_3$ to $S_1$},
%\\\no
%%
%&& \text{move}~\{y_2y^2, y_1y^2, y^3, y^2\}~\text{in $S_4$ to
%$S_3$},\\\no
%%
%&& \text{move}~\{y_2y_1y, y_2y_1\}~ \text{in $S_4$ to $S_1$}
%\end{eqnarray}
%Obviously,  the rearranged matrix $\widehat{D}_{2, 2}$ is also
%square and not equal to zero identically by Corollary 3.3. As will
%be demonstrated in Subsection \ref{sec-s2},  the matrix
%$\widehat{D}_{2, 2}$ is identical to some algebraic sparse resultant
%matrix of the system $\{g_1, g_2, \delta g_1, \delta g_2\}$ regarded
%as the polynomials of variables $y, y_1, y_2$.
%*****Can we know the exact degree for the resultant in this case?

\section{Conclusion and discussion}
In this paper,  a matrix representation for two first order
nonlinear generic ordinary differential polynomials $f_1, f_2$ is
given. That is,  a non-singular matrix is constructed such that its
determinant contains the differential resultant as a factor.  The
constructed matrix is further shown to be an algebraic sparse matrix
of $f_1, f_2, \delta f_1, \delta f_2$ when certain special lift
functions are used. Combining the two results,  we show that the
sparse resultant of $f_1, f_2, \delta f_1, \delta f_2$ is not zero
and conatins the differential resultant of $f_1, f_2$ as a factor.

It can be seen that to give a matrix representation for $n+1$
generic polynomials in $n$ variables is far from solved,  even in
the case of $n=1$. Based on what is proved in this paper,  we
propose the following conjecture.

\textbf{Conjecture}. Let $\mathcal {P}=\{f_1, f_1, \dots, f_{n+1}\}$
be $n+1$ generic differential polynomials in $n$ indeterminates,
$\mbox{ord}(f_i)=s_i$, and $s=\sum_{i=0}^n s_i$.
Then the sparse resultant of the algebraic polynomial system
\begin{eqnarray}\label{eq-gas}
f_1, \delta f_1, \dots\delta^{s-s_0}f_1, \ldots, f_{n+1}, \delta
f_{n+1}, \dots\delta^{s-s_n}f_{n+1}
\end{eqnarray}
is not zero and contains the differential resultant of $\mathcal
{P}$ as a factor.
%
%Furthermore,  for certain lift functions, the sparse resultant
%matrix for \bref{eq-gas} is nonsingular.

%As an exploration,  we consider differential resultant of two
%differential polynomials with degree two in which one is order two
%and another is order one.
%
%Suppose $\text{ord}(f_1)=2, \text{ord}(f_2)=1,
%\text{deg}(f_1)=\text{deg}(f_2)=2$,  choose $l_1=(3,  -2,  1, 1),
%l_2=(1,  2,  1, 1),  l_3=(2,  3,  -1, 2),  l_4=(-1,  2, 3, 1),
%l_5=(2,  -1,  -3, 5)$,  which correspond to lift functions of
%polytope of $\{f_1, f_2, \delta f_1, $ $\delta f_2, \delta^2 f_2\}$
%respectively,  then we obtain their main monomials are
%$\mbox{mm}(f_1)= y_1^2$,  $\mbox{mm}(f_2)=1$,  $\mbox{mm}(\delta
%f_1)=y_3y_2$,  $\mbox{mm}(\delta f_2)=y^2$,  $\mbox{mm}(\delta^2
%f_2)=y_2^2$. Following the method as Lemma \ref{lm-32},  one can
%prove the determinant of the sparse resultant matrix of the system
%$\{f_1, f_2, \delta f_1, $ $\delta f_2, \delta^2 f_2\}$ is not equal
%to zero identically. Moreover,  similar as Theorem \ref{th-2},  the
%differential resultant of $f_1$ and $f_2$ is a factor of the
%determinant.

\section*{Acknowledgments}
This article is partially supported by a National Key Basic Research
Project of China (2011CB302400) and by grants from NSFC (60821002,
11101411). Z.Y. Zhang  acknowledges the support of  China
Postdoctoral Science Foundation funded project (20110490620).


\begin{thebibliography}{99}
%\bibitem{bez-1779} E. B$\acute{\text{e}}$zout,
% Th$\acute{\text{e}}$orie G$\acute{\text{e}}$n$\acute{\text{e}}$rale  des  $\acute{\text{E}}$quations  Alg$\acute{\text{e}}$briques.  Paris,
%1779.

\bibitem{berkovich}
 L.M. Berkovich and V.G. Tsirulik.
 Differential resultants and some of their applications.
 Differ. Equation. 22 (1986) 750-757.

%\bibitem{bs}
% L. Billera and B. Sturmfels.
% Fiber polytopes. {\em Ann. of Math.},  135 (1992),  527-549.

%\bibitem{canny1}
% J.~F. Canny.
% Generalized Characteristic Polynomials.
%  {\em Journal of Symbolic Computation},  9,  241-250,  1990.

\bibitem{ce1}
 J.F.  Canny and  I.Z. Emiris.
 Efficient Incremental Algorithms for the Sparse Resultant and the Mixed
 Volume.
 {\em Journal of Symbolic Computation},
 20(2),  (1995),  117-149.
%
% An  Efficient Algorithm for the Sparse Mixed Resultant.
% In  {\em Proc. Int. Symp. on  Appl. Algebra,  Algebraic Algorithms and Error-Corr.
% Codes},  Puerto Rico,  LNCS 263 (1993) 89-104.

\bibitem{ce2}
 J.F. Canny and  I.Z. Emiris,
 A Subdivision-based  Algorithm  for  the Sparse Resultant.
 J. ACM  47(3)(2000) 417-451.

\bibitem{dres1}
 G. Carr\`a-Ferro.
 A Resultant Theory for the Systems of Two Ordinary Algebraic Differential Equations.
 {\em Applicable Algebra in Engineering,  Communication and Computing},  8,  539-560,  1997.

\bibitem{chardin1}
 M. Chardin.
 Differential Resultants and Subresultants.
 {\em Fundamentals of Computation Theory},  LNCS,  Vol. 529,  180-189,
 Springer-Verlag,   Berlin,  1991.

\bibitem{cox-1998}
 D. Cox,  J. Little and D. O'Shea.
 Using Algebraic Geometry,  Springer-Verlag,  New York,  1998.

\bibitem{dandrea1}
 C. D'Andrea.
 Macaulay style formulas for the sparse resultant.
 {\em Trans. Amer. Math. Soc.} 354 (2002) 2595-2629.

\bibitem{gao}
 X.S. Gao,  W. Li, and C.M. Yuan.
 Intersection theory in differential algebraic geometry:
 generic intersections and the differential Chow form.
 Accepted by {\em Trans. Amer. Math. Soc.}, 58 pages. Also in arXiv:1009.0148v2.

\bibitem{lp}
 S.I. Gass.
 Linear Programming: Methods and applications, (5th ed.)
 McGraw-Hill,  Inc. New York, USA, 1984.

\bibitem{gel}
 I.M. Gelfand,  M. Kapranov, and A. V. Zelevinsky.
 Discriminants,  Resultants and Multidimensional Determinants.
 Boston,  Birkh¨auser,  1994.

\bibitem{hhong}
 H. Hong.
 Ore subresultant coefficients in solutions.
 {\em Applicable Algebra in Engineering,  Communication and Computing},
 12(5),  421-428,  2001.

\bibitem{liw1}
 W. Li,  X.S. Gao, and C.M. Yuan.
 Sparse differential resultant.
 In Proc. ISSAC 2011,  San Jose,  CA,  USA,  225-232,  ACM Press,  New York,  2011.

\bibitem{liw2}
 W. Li, C.M. Yuan, and X.S. Gao.
 Sparse differential resultant for Laurent differential polynomials.
 arXiv:1111.1084v2, 2012.

\bibitem{zmli}
 Z. Li.
 A Subresultant Theory for Linear Differential,  Linear Difference and Ore Polynomials,
 with Applications.
 PhD thesis,  Johannes Kepler University,  1996.

\bibitem{Ma-1916}
 F.S. Macaulay.
 The Algebraic Theory of Modular Systems. Cambridge: Proc. Cambridge Univ. Press 1916.

%\bibitem{Ma-1903}
% F.S. Macaulay.
% Some Formulae in Elimination. Proc. Lond. Math. Soc. 35 (1903) 3-27.

\bibitem{stu1}
 B. Sturmfels.
 Sparse elimination theory.
 In D. Eisenbud and  L. Robbiano,   editors,  Computation Algebraic
 Geometry and Commutative  Algebra,  Proceedings,  Cortona,  June
 1991,
 Cambridge  University  Press,  264-298.

\bibitem{stu2}
 B. Sturmfels.
 On the Newton polytope of the resultant.
 {\em Journal of Algebraic Combinatorics}, 3, 207-236, 1994.

%\bibitem{renegar1}
% J. Renegar.
% On the Computational Complexity and Geometry of the First-order
% Theory of the Reals,  Part I.
% %: Introduction. Preliminaries.
% %The Geometry of Semi-algebraic Sets.
% %The Decision Problem for the Existential Theory of the Reals.
% {\em Journal of Symbolic Computation},  13(3),  255-299,  1992.

\bibitem{ritt0}
 J.~F. Ritt.
 {\em Differential Equations from the Algebraic Standpoint}.  Amer. Math. Soc.,   New York,  1932.

\bibitem{lres2}
 S.~L. Rueda.
 Linear sparse differential resultant formulas.
 arXiv:1112.3921v1,  Dec. 2011.

\bibitem{lres1}
 S.~L. Rueda and J.~R. Sendra.
 Linear Complete Differential Resultants and the Implicitization of Linear DPPEs.
 {\em Journal of Symbolic Computation},  45(3),  324-341,  2010.


%\bibitem{syl-1853} J.J.  Sylvester,  On  a  Theory  of  Syzygetic  Relations  of  Two  Rational  Integral  Functions,
%Comprising  an  Application  to  the  Theory  of Sturm's  Functions,
%and  that  of the  Greatest Algebraic  Common  Measure.
%Philosophical   Trans.  143 (1853) 407-548.

\bibitem{yang-dixon}
 L. Yang,  Z. Zeng and  W. Zhang.
 Differential Elimination with Dixon Resultants.
 Abstract in {\em Proc. ACA2009},  June 2009,  Montreal,  Canada.
 Preprint No.~2,  SKLTC,  East China Normal University,  July,  2011.

\bibitem{handbook}
 D. Zwillinger.
 {\em Handbook of Differential Equations}.
 Academic Press,  San Diego,  USA,  1998.


\end{thebibliography}
\end{document}